\documentclass[
superscriptaddress,
twocolumn,
amsmath,amssymb,
aps,
prb,
]{revtex4-2}

\usepackage{graphicx,framed} 
\usepackage{dcolumn} 
\usepackage{bm} 
\usepackage{braket}
\usepackage{dsfont}
\usepackage{color,xcolor}
\usepackage{amsthm}
\usepackage{hyperref} 
\usepackage[normalem]{ulem} 
\usepackage{qcircuit}       
\usepackage[export]{adjustbox}

\usepackage{amsthm}

\usepackage{comment}

\usepackage{algorithm}
\usepackage[noend]{algpseudocode}
\algdef{SE}[DOWHILE]{Do}{doWhile}{\algorithmicdo}[1]{\algorithmicwhile\ #1}
\makeatletter
\def\algbackskip{\hskip-\ALG@thistlm}
\makeatother

\definecolor{lightblue}{RGB}{73,151,208}
\definecolor{crimson}{RGB}{140,41,53}

\hypersetup{
    colorlinks,
    linkcolor={crimson},
    citecolor={lightblue},
    urlcolor={lightblue}
}

\theoremstyle{definition}

\newcommand{\tr}{\mathrm{Tr}}

\newcommand{\mD}{\mathcal{D}}

\newcommand{\mU}{\mathcal{U}}

\newcommand{\mE}{\mathcal{E}}

\newcommand{\mK}{\mathcal{K}}
\newcommand{\mT}{\mathcal{T}}

\newcommand{\Lbeta}{L_{\beta}}
\newcommand{\p}{p}
\newcommand{\q}{q}

\newcommand{\sgn}{\mathrm{sgn}}
\newcommand{\lset}{\{{\p}_i\}}
\newcommand{\qdet}[1]{\text{Qdet}{\left[#1\right]}}
\begin{document}

\preprint{}

\title{Exact spectral form factors of non-interacting fermions with Dyson statistics}

\author{Tatsuhiko N. Ikeda}
\email{tatsuhiko.ikeda@riken.jp}
\affiliation{Department of Physics, Boston University, Boston, Massachusetts 02215, USA}
\affiliation{RIKEN Center for Quantum Computing, Wako, Saitama 351-0198, Japan}

\author{Lev Vidmar}
\email{lev.vidmar@ijs.si}
\affiliation{Department of Theoretical Physics, J. Stefan Institute, SI-1000 Ljubljana, Slovenia}
\affiliation{Department of Physics, Faculty of Mathematics and Physics, University of Ljubljana, SI-1000 Ljubljana, Slovenia}

\author{Michael O. Flynn}
\email{moflynn@bu.edu}
\affiliation{Department of Physics, Boston University, Boston, Massachusetts 02215, USA}
\affiliation{Department of Physics \& Astronomy, University of Victoria, Victoria, British Columbia V8P 5C2, Canada}

\begin{abstract}
The spectral form factor (SFF) is a powerful diagnostic of random matrix behavior in quantum many-body systems. We introduce a family of random circuit ensembles whose SFFs can be computed \textit{exactly}. These ensembles describe the evolution of non-interacting fermions in the presence of correlated on-site potentials drawn from the eigenvalue distribution of a circular ensemble. For disorder parameters drawn from the circular unitary ensemble (CUE), we derive an exact closed form for the SFF, valid for any choice of system size $L$ and integer time $t$. When the disorder is drawn from the circular orthogonal or symplectic ensembles (COE and CSE, respectively), we carry out the disorder averages analytically and reduce the computation of the SFF at integer times to a combinatorial problem amenable to transfer matrix methods. In each of these cases the SFF grows exponentially in time, which we argue is a signature of random matrix universality at the single-particle level. Finally, we develop matchgate circuit representations of our circuit ensembles, enabling their experimental realization in quantum simulators.
\end{abstract}

\maketitle
\tableofcontents

\section{Introduction}\label{sec:Intro}
In the era of precision microscopic experiments, physicists have been forced to contend with foundational questions in statistical mechanics and many-body dynamics~\cite{eisert2015quantum, ETHReview, abanin2019colloquium,ThermalizationOverview, sierant_lewenstein_24}. On the experimental side, quantum simulators~\cite{RMP_Simulators,monroe2021programmable} and ultracold atomic experiments~\cite{BlochQuantumSimulators,QuantumSimulatorReview,langen2015ultracold,Science_Simulators} have offered unprecedented insight into the processes which lead to thermalization. In parallel, theorists have developed a framework for studying dynamical properties of quantum systems, including the eigenstate thermalization hypothesis (ETH), and put forward practical diagnostics of chaos~\cite{DeutschStatMech, Srednicki1, Srednicki2, Srenicki3, rigol_dunjko_08,AdiabaticEigenstateDeformations}.

An important theoretical tool for analyzing quantum dynamics is the spectral form factor (SFF), introduced in pioneering work by Berry~\cite{BerrySpectralRigidity} and defined as follows. Given an ensemble $\mE$ of time-evolution operators, the SFF is defined by
\begin{align}\label{eq:SFFDef}
K_{\mE}(t) = \langle|\tr\left(U^t\right)|^{2}\rangle_{\mE},
\end{align}
where $U\in\mE$, $\langle\cdot\rangle_{\mE}$ denotes ensemble averaging and $t$ parameterizes continuous (discrete) time for Hamiltonian (Floquet) evolution. The SFF encodes fine-grained dynamical properties of the ensemble, which makes it a useful tool to distinguish integrable and chaotic systems in accordance with the Berry--Tabor~\cite{BerryTabor} and Bohigas--Giannoni--Schmidt~\cite{BGS} conjectures.

The SFF is qualitatively well-understood in a number of important cases, including time-evolution operators drawn from a circular ensemble or generated by a Hamiltonian drawn from a Gaussian ensemble. These models effectively describe a large class of interacting quantum systems, where the symmetries of the system determine the relevant ensemble. In these cases, the SFF tends to exhibit the so-called ``dip-ramp-plateau'' structure, and it is widely accepted that a linear ramp in the SFF, $K_{\mE}(t)\sim t$, serves as a signature of random matrix universality~\cite{Kos18, Bertini2018, ChanMinimalModel, Chan2018, Liu2018, ChanConservedCharge, suntajs_bonca_20a, sierant_delande_20, suntajs_prosen_21, Vasilyev20, Prakash21, Dibyendu1, Dibyendu2, Liao2, ErgodicityBreakingZeroDimensions, Joshi22, HydrodynamicSFF, Liu17, Gharibyan18,  DissipativeSFF, BrokenUnitaritySFF, Dag2023}. Other dynamical regimes, such as hydrodynamics, have also been identified through the SFF~\cite{HydrodynamicSFF}.

In contrast, a number of random matrix ensembles have recently been identified whose form factors do not exhibit a linear ramp. Examples include ensembles of Clifford circuits~\cite{Farshi2023} and systems of non-interacting fermions with single-particle states that exhibit random matrix statistics~\cite{Winer2020,Liao2020}. In these cases, the SFF grows rapidly; indeed, numerical calculations are more consistent with an exponential ramp rather than a linear one. Unfortunately, carrying out a rigorous analysis of the SFF is difficult in general. Numerically, the SFF is not self-averaging and convergence of the disorder average in~\eqref{eq:SFFDef} requires a tremendous number of samples~\cite{Prange1997}. Analytical treatments of the SFF also tend to rely on simplifying properties of large $N$ limits and may employ poorly controlled approximations. Given the prominence of the SFF as a probe of quantum dynamics, identifying models which qualitatively deviate from the linear ramp and can be rigorously analyzed is an important challenge in current research.

In this work, we investigate the dynamics of non-interacting fermions in the presence of spatially correlated potentials. We will show that when these potentials are drawn from the eigenvalue distribution of a circular ensemble, the resulting \textit{many-body} SFF can be computed exactly. More precisely, let $U_{\beta}$ denote a random $L_{\beta}\times L_{\beta}$ unitary drawn from the circular ensemble with Dyson index $\beta$, where $\beta=1,2,4$ denote the circular orthogonal, unitary, and symplectic ensembles (COE, CUE, and CSE, respectively), and
\begin{align}
    L_\beta \equiv \begin{cases}
        L & (\beta=1,2)\\
        2L & (\beta=4).
    \end{cases}
\end{align}
The eigenvalues of $U_{\beta}$ can be written as $e^{i\theta_j}$, where $\theta_j\in [-\pi,\pi)$ are the quasienergies (for a discussion of gauge fixing, see Sec.~\ref{ssec:SingleParticleSFFs}). By construction, the quasienergies exhibit random matrix statistics which define the single-particle sector of our model. In particular, the quasienergies define a many-body unitary $\mU_{\beta}$ for an $L_{\beta}$-site fermion system,
\begin{equation}\label{eq:ManyBodyFloquet}
    \mU_{\beta}=\exp\left(-i\sum_{j=1}^{L_{\beta}}\theta_{j}n_{j}\right),
\end{equation}
where $n_{j}$ are fermion number operators and $\mU_{\beta}$ has dimension $\mD_{\beta}=2^{L_{\beta}}$. We will focus on the many-body SFFs of these ensembles, \begin{align}\label{eq:MB_SFF}
\mK_{\beta}(t) = \left\langle|\tr\left(\mU_\beta^t\right)|^{2}\right\rangle,
\end{align}
where the ensemble average is implicitly understood to be over the appropriate distribution of quasienergies. We refer to the ensembles of unitaries defined by~\eqref{eq:ManyBodyFloquet} as the \textit{single-particle} circular ensembles, in contrast to the conventional circular ensembles defined on the $\mD_\beta$-dimensional Hilbert space.

As announced in the companion paper~\cite{OurPaper1}, we have derived methods to compute the spectral form factors of the single-particle circular ensembles exactly. For the single-particle CUE, we have derived an exact analytical result for $\mK_1(t)$ which holds for arbitrary choices of $L$ and integer $t$. For the single-particle COE and CSE, we have performed the ensemble averages analytically and reduced the computation of the SFFs to combinatorial problems that can be solved exactly by transfer matrix methods when $t$ is an integer. Related methods can be used to extract the SFFs for arbitrary values of $t$ but are not amenable to closed-form expressions in general. In each case, the SFF exhibits exponential growth in time. The exponential growth occurs naturally in non-interacting systems which nonetheless exhibit level repulsion due to the choice of disorder. This article contains full derivations of these previously announced results in addition to developing circuit realizations of the single-particle circular ensembles with matchgates. Our findings complement the existing understanding of spectral statistics in interacting systems~\cite{Bertini2018,Chan2018} and demonstrates that exponential growth of the form factor is a signature of random matrix universality at the single-particle level~\cite{Winer2020,Liao2020}.

The remainder of the paper is organized as follows. In Sec.~\ref{sec:formulation}, we review well-known results for the circular ensembles and introduce the moment expansion as a method of computing the many-body SFF~\eqref{eq:MB_SFF}. The formalism of this section applies equally to each of the single-particle circular ensembles; ensemble-specific considerations are relegated to later sections.
In Sec.~\ref{sec:CUE}, we study $\mK_2(L, t)$ and map its computation onto a combinatorial problem, which we solve to obtain the exact SFF for any $L$ and integer $t$. In addition, we present methods to compute $\mK_2(t)$ for arbitrary real $t$ and derive simple closed-forms that describe the short and long-time limits.
In Sec.~\ref{sec:CSEandCOE}, we analyze the moment expansions of the COE and CSE and develop transfer matrix methods that compute their form factors exactly. 
In Sec.~\ref{sec:circuits}, we introduce ensembles of matchgate unitaries which reproduce the form factors of Secs.~\ref{sec:CUE} and~\ref{sec:CSEandCOE}, providing a path for implementing such ensembles in quantum simulators. Finally, we conclude and discuss future directions in Section~\ref{sec:conclusion}.

\section{Overview of single-particle circular ensembles}\label{sec:formulation}
In this section, we review important features of the circular ensembles and establish a general formalism to compute the spectral form factors of the single-particle circular ensembles. Subsequent ensemble-specific sections will establish additional formalism as needed.

\subsection{Review of the circular ensembles}
 For our purposes, the most important random matrix ensembles are the circular ensembles introduced by Dyson. The spectral form factors of these ensembles are analytically tractable and we briefly review some of their essential properties here; more complete reviews can be found in Refs.~\cite{mehta1991random,DysonEigenvalueCorrelations}. 
 
Let $U_{\beta}$ be a random unitary as defined in Sec.~\ref{sec:Intro} with quasienergies $\theta_j$. The $\theta_j$ are governed by the joint probability distributions~\cite{mehta1991random}
\begin{align}\label{eq:CircularDistributions}
    P_{\beta}(\theta_{1},\cdots,\theta_{L}) &= \frac{1}{Z_{L,\beta}}\prod_{j<k}|e^{i\theta_{j}}-e^{i\theta_{k}}|^{\beta},\\ Z_{L,\beta}&=\left(2\pi\right)^{L}\frac{\Gamma(\beta L/2 +1)}{\left(\Gamma(\beta/2+1)\right)^{L}}.    
\end{align}
The attentive reader will notice that for the CSE ($\beta=4$), the random unitary $U_4$ has dimension $L_{4}=2L$ while only $L$ quasienergies are defined in~\eqref{eq:CircularDistributions}. This is accounted for by noting that the CSE quasienergy spectrum is two-fold degenerate for each disorder realization.

Using these distributions, many properties of the circular ensembles can be computed analytically. In particular, the SFFs of the circular ensembles~\eqref{eq:SFFDef} are known~\footnote{Strictly speaking, the formulae for $K_1(t)$ and $K_4(t)$ are exact only in the limit $L\to\infty$, while~\eqref{eq:SingleParticleSFF2} does not suffer from finite size effects and holds for arbitrary $L$.}:
\begin{align}
K_{1}(t) &=\begin{cases}\label{eq:SingleParticleSFF1}
2t-t\log\left(1+\frac{2t}{L}\right) & (0<t\leq L)\\
2L-t\log\left(\frac{2t/L +1}{2t/L -1}\right) & (t\geq L),
\end{cases}\\
K_{2}(t) &=\begin{cases}\label{eq:SingleParticleSFF2}
t & (0<t<L)\\
L & (t\geq L),
\end{cases}\\
K_{4}(t) &=\begin{cases}\label{eq:SingleParticleSFF3}
2t-t\log|1-t/L| & (0<t\leq 2L)\\
4L & (t>2L).
\end{cases}
\end{align}
\begin{figure}
    \centering
    \includegraphics[width=\columnwidth]{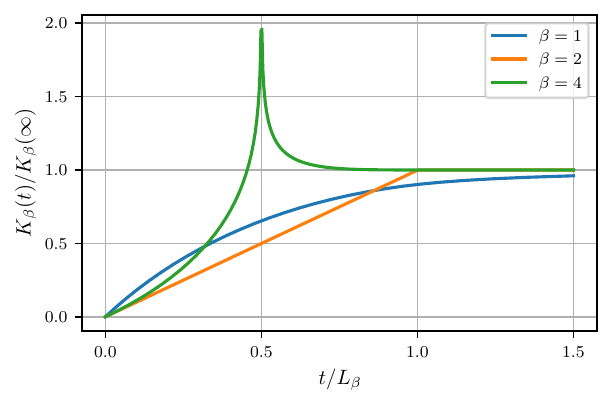}
    \caption{Normalized SFFs~\eqref{eq:SFFDef} of Dyson's circular ensembles. The emergence of a late-time ramp is often taken as a defining feature of random matrix universality in quantum systems.
    }
    \label{fig:singleparticleSFF}
\end{figure}
Normalized plots of the SFF are shown in Fig.~\ref{fig:singleparticleSFF}. A particularly important feature is the approximate $t$-linear behavior for $\beta=1,2$, which is often referred to as the linear ramp. In Hamiltonian systems, convergence of the SFF to one of the Dyson results is usually taken as a clear signature of random matrix universality and the symmetries of the system determine the appropriate circular ensemble.

To close this section, we note that the form factors of Eqs.~\eqref{eq:SingleParticleSFF1}--\eqref{eq:SingleParticleSFF3} hold for circular  ensembles of dimension $\mD_{\beta}$ via the replacement $L_{\beta}\to\mD_{\beta}$. In particular, this implies that the timescale at which the SFF saturates -- the (many-body) Heisenberg time -- scales with the Hilbert space dimension, $t_H\sim\mD$.

\subsection{SFF of the single-particle circular ensembles}\label{ssec:SingleParticleSFFs}
The many-body unitaries~\eqref{eq:ManyBodyFloquet} introduced in Sec.~\ref{sec:Intro} are naturally understood in terms of non-interacting fermions. In particular, the eigenvalues of $\mU_{\beta}$ can be written in the form $e^{i\Theta}$, where the many-body quasienergies $\Theta$ are specified by the fermion occupation numbers
\begin{equation}\label{eq:Ueval}  \Theta(\bm{n}) = \sum_{i=1}^{L_{\beta}} \theta_{i}n_{i}.
\end{equation}
Here we have introduced a convention that bold text denotes a vector. When the dimension of a vector is unclear, we specify it with an explicit subscript, e.g., $\bm{p}_{n}=\left(p_{1},\cdots,p_{n}\right)$.

The many-body SFFs of the single-particle circular ensembles are given by
\begin{align}
    \mK_\beta(t) &= \left\langle\left|\tr\; \left(\mU_{\beta}^{t}\right)\right|^2\right\rangle\\
    &= \left\langle\left| \prod_{j=1}^{L_{\beta}}\sum_{n_{j}=0}^1 e^{-i\theta_j n_j t} \right|^2 \right\rangle\\
    &= 2^{L_{\beta}} \int d\bm{\theta} P_\beta(\bm{\theta}) \prod_{j=1}^{L}\left[1+C_{\beta}\left(t\theta_{j}\right)\right],\label{eq:SFF}
\end{align}
where
\begin{align}\label{eq:Cfunc}
    C_{\beta}(x) \equiv \begin{cases}
        \cos(x) & (\beta=1,2)\\
        2\cos(x)+\cos^2(x) & (\beta=4).
    \end{cases}
\end{align}
We remind the reader that the quasisenergy distribution of the CSE ($\beta=4$) is two-fold degenerate, which is essential to the derivation of $C_4(x)$.

The SFFs $\mK_{\beta}(t)$ depend, in general, on the range in which the quasienergies $\theta_j$ are defined. Due to the obvious similarities with other gauge structures, we refer to transformations which map $\bm{\theta}\to\bm{\theta}+2\pi\bm{m}, m_{j}\in\mathbb{Z},$ as gauge transformations. In this language, $\mK_{\beta}(t)$ is gauge invariant if and only if $t$ is an integer and the choice $\theta_j\in\left[-\pi,\pi\right)$ is an example of gauge fixing. Despite our use of this terminology, the reader should keep in mind that the ``gauge choice'' carries physical meaning: for example, shifting $\theta_j$ in the free fermion evolution operator~\eqref{eq:ManyBodyFloquet} amounts to shifting a chemical potential, which has physical consequences.

As a final comment, we note that the joint distribution function $P_\beta(\bm{\theta})$ contains \textit{all} correlations among the quasienergies, such as level repulsion. At the single-particle level, these correlations encode the spectral form factors of equations~\eqref{eq:SingleParticleSFF1}--\eqref{eq:SingleParticleSFF3}; in the many-body context, they describe exponential growth of the functions $\mK_{\beta}(t)$.

\subsection{Moment expansion of the SFF}\label{sec:SFFfromMoments}
In this section, we relate the many-body SFFs $\mK_{\beta}(t)$ to $n$-point functions of the circular ensembles. The $n$-point functions are well-known in the random matrix theory literature~\cite{mehta1991random} and we will not review them in detail here.

Using the fact that $P_\beta(\bm{\theta})$ is invariant under permutations of the $\{\theta_{i}\}$, the SFF~\eqref{eq:SFF} can be organized by the number of variables with a non-trivial integrand,
\begin{equation}\label{eq:SingleParticleExpansion}
\begin{aligned}
    &\mK_\beta(t) = 2^{L_{\beta}}\int d\bm{\theta} P_\beta(\bm{\theta})\sum_{n=0}^{L}\;\; \binom{L}{n}\prod_{j=1}^{n}C_{\beta}\left(t\theta_{j}\right)\\
    &= 2^{L_{\beta}}\left[1+\sum_{n=1}^{L}\int\left(\prod_{i=1}^{n}C_{\beta}\left(t\theta_{i}\right)\;d\theta_{i}\right)\frac{\mathtt{R}^\beta_{n}(L,\bm{\theta}_{n})}{n!}\right],
\end{aligned}
\end{equation}
where $\mathtt{R}_{n}^\beta$ is the $n$-point single-particle correlation function with Dyson index $\beta$, defined as
\begin{equation}
    \mathtt{R}^\beta_{n}(L,\bm{\theta}_{n}) = \frac{L!}{(L-n)!}\int d\theta_{n+1}\cdots d\theta_{L}\;\; P_\beta(\bm{\theta}).
\end{equation}
The SFF is then reduced to the sum
\begin{align}
    \mK_\beta(t)
    &= 2^{\Lbeta}\left[1+\sum_{n=1}^{L}\frac{\mathtt{r}^\beta_{n}(L,t)}{n!}\right],\label{eq:SFFMomentExpansion}\\
    \mathtt{r}^\beta_{n}(L,t) &\equiv     \int d\bm{\theta}_{n}\; \mathtt{R}^\beta_{n}(\bm{\theta}_{n})\; \prod_{i=1}^{n}C_{\beta}\left(t\theta_{i}\right).\label{eq:DefMoments}
\end{align}
We refer to Eq.~\eqref{eq:SFFMomentExpansion} as a moment expansion and to $\mathtt{r}_{n}^{\beta}(L,t)$ as the $n$th moment. The moment expansion is named in analogy with the cumulant expansion of Ref.~\cite{Liao2020}, which developed a similar picture to our own for the Gaussian unitary ensemble (GUE). An important technical distinction between the circular and Gaussian ensembles is encoded in their eigenvalues: while the quasienergies $\theta_{i}$ lie in a finite range, the eigenvalues of GUE matrices are unbounded on the real line. This distinction allows us to avoid introducing approximations when evaluating disorder averages for the circular ensembles, such as the box approximation for Gaussian ensembles~\cite{Cotler2017,Liu2018,Liao2020}.

\section{Single-particle CUE $(\beta=2)$}\label{sec:CUE}
In this section, we present an exact calculation for the SFF of the single-particle circular unitary ensemble, $\mK_{2}(t)$. As this is the only ensemble considered in this section, we leave out explicit ensemble labels; for example, we will drop the Dyson index on the many-body SFF and write $\mK(t)$. When $t$ is an integer, our approach yields a simple closed form for the SFF in any choice of system size; for arbitrary real $t$, our methods are still exact but cannot be evaluated in closed form.

Our strategy for evaluating the SFF is as follows. First, we consider the case where $t$ is an integer and show in Sec.~\ref{sec:MomentsFromDimers} that the computation of the moments can be mapped onto a combinatorial problem. Using this representation, in Sec.~\ref{sec:Factorization} we derive a factorization identity for the SFF which allows us to compute the SFF for any system size $L$ and integer time $t$. We then present our final result for integer $t$ in Sec.~\ref{ssec:ExactCUE} and discuss the generalization to arbitrary $t$ in Sec.~\ref{ssec:RealtimeMoments}.

\subsection{Computing moments with dimer embeddings}\label{sec:MomentsFromDimers}

To proceed with the moment expansion from Eq.~\eqref{eq:SFFMomentExpansion}, we will derive an explicit form for the moments $\mathtt{r}_{n}(L,t)$ when $t$ is an integer, which we assume throughout this section. First, expand $C(t\theta_i)=\cos(t\theta_i)$ in exponentials so that Eq.~\eqref{eq:DefMoments} becomes
\begin{align}
    \mathtt{r}_{n}(L,t) &\equiv     \frac{1}{2^{n}}\sum_{\bm{\xi}_{n}}\int d\bm{\theta}_{n}\; \mathtt{R}_{n}(\bm{\theta}_{n})\;e^{it\bm{\theta}_{n}\cdot\bm{\xi}_{n}},\label{eq:CUEMoments}
\end{align}
where $\bm{\xi}_{n}=\left(\xi_{1},\cdots,\xi_{n}\right)$ with $\xi_{i}=\pm 1$.
To further simplify the moments, it is convenient to replace the single-particle correlation functions with the CUE kernel, $\mathtt{K}$, defined as~\cite{mehta1991random}
\begin{align}\label{eq:KernelCUE}
    \mathtt{K}(\theta_i,\theta_j) =
    \begin{cases}
        \mathtt{R}_1(\theta_i) =\frac{L}{2\pi} & (i=j)\\
        \mathtt{K}(\theta_i-\theta_j) = \frac{\sin \frac{L}{2}(\theta_i-\theta_j)}{2\pi \sin \frac{1}{2}(\theta_i-\theta_j)}   & (i\neq j)\,.
    \end{cases}
\end{align}
In terms of the CUE kernel, the single-particle correlation functions are given by the determinant of the kernel matrix
\begin{equation}\label{eq:rn_det}
\begin{aligned}
\mathtt{R}_{n}(\bm{\theta}) &= \text{det}\left[\mathtt{K}(\theta_{j}-\theta_{k})\right]_{j,k = 1,\cdots, n} \\
    &=\sum_{\sigma\in\mathbb{S}_n}\sgn(\sigma)\prod_{j=1}^{n}\mathtt{K}(\theta_j-\theta_{\sigma(j)}),
\end{aligned}
\end{equation}
where $\mathbb{S}_{n}$ is the symmetric group of order $n$ and $\sgn(\sigma)$ is given by 1 ($-1$) if $\sigma$ is an even (odd) permutation.
The kernel function~\eqref{eq:KernelCUE} admits a convenient Fourier series representation, 
\begin{align}\label{eq:KFourier}
    \mathtt{K}(\theta)  &= \frac{1}{2\pi}e^{-i\frac{L-1}{2}\theta}\sum_{{\p}=0}^{L-1}e^{i{\p} \theta}.
\end{align}
Plugging Eqs.~\eqref{eq:rn_det} and \eqref{eq:KFourier} into Eq.~\eqref{eq:DefMoments}, we find
\begin{widetext}
\begin{align}\label{eq:MomentSums}
    \mathtt{r}_n(L,t) 
    &= \frac{1}{2^n}\sum_{\sigma\in\mathbb{S}_n}\sgn(\sigma)\sum_{{\p}_1,\dots,{\p}_n=0}^{L-1}\sum_{\bm{\xi}_{n}}\int\frac{d\bm{\theta}_{n}}{(2\pi)^n}\exp\left[i\sum_{j=1}^n ({\p}_j-{\p}_{\sigma(j)}+t\xi_j )\theta_j\right]\\
    &= \frac{1}{2^n}\sum_{\sigma\in\mathbb{S}_n}\sgn(\sigma)\sum_{{\p}_1,\dots,{\p}_n=0}^{L-1}\sum_{\bm{\xi}_{n}}\prod_{j=1}^{n} \delta({\p}_j-{\p}_{\sigma(j)}+t\xi_j),
\end{align}
\end{widetext}
where we have used continuum notation for the Kronecker delta, i.e.,
$\delta (m) \equiv \delta_{m,0}$.

We note that $\mathtt{r}_{n}(L,t>0)$ vanishes whenever $n$ is odd due to the following argument. The terms which contribute to $\mathtt{r}_{n}$ necessarily satisfy ${\p}_{j}-{\p}_{\sigma(j)}+t\xi_{j}=0$ for all $j\in\{1,\dots,n\}$, which implies that $t\sum_{j}\xi_{j} = 0$. For any $t\neq 0$, we then require that $\sum_{j}\xi_{j}=0$, which cannot be the case when $n$ is odd. Going forward, we will work exclusively with the even moments $\mathtt{r}_{2n}$ and assume $t>0$.

\begin{figure}
    \centering
    \includegraphics[width=\columnwidth]{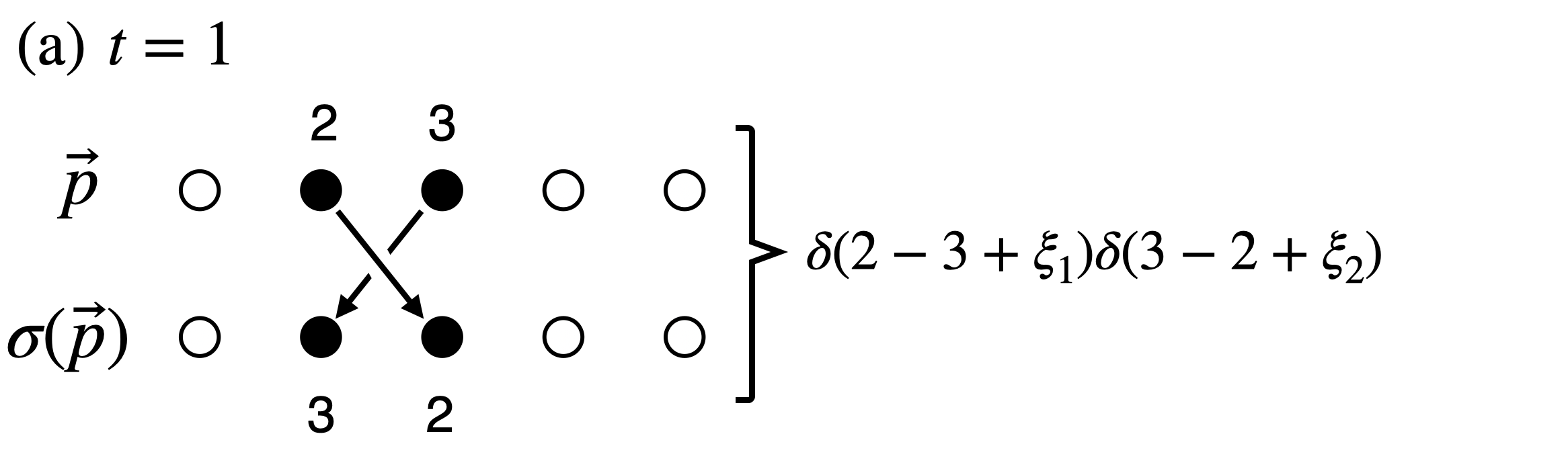}\\
    \vspace{5mm}
    \includegraphics[width=\columnwidth]{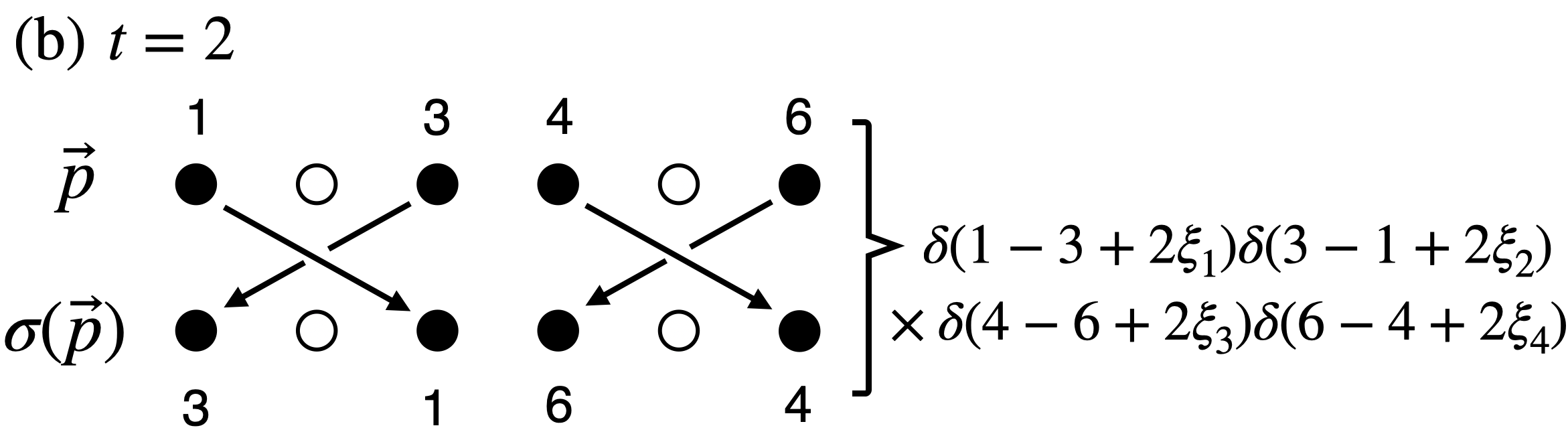}
    \caption{Pictorial representation of terms which contribute to $Q_{n}(L,t)$. Solid (empty) circles indicate occupied (unoccupied) sites. (a) A contribution to $Q_{1}(L=5,t=1)$. Note that there is a unique choice of $\sigma$ and $\bm{\xi}_{2}$ which contribute to $Q_{1}(5,1)$. We say that the particles on sites 2 and 3 form a dimer. (b) A contribution to $Q_{2}(L=6,t=2)$. Again, $\sigma$ and $\bm{\xi}_{4}$ are uniquely specified by $\lset$ and in this case the particles on sites (1,3) and (4,6) form range-2 dimers.}
    \label{fig:cue_pic1}
\end{figure}

To evaluate the even moments, we note that the sums which contribute to $\mathtt{r}_{2n}$ are invariant under permutations of the $\lset$. Furthermore, terms with ${\p}_{i}={\p}_{j}$ for some $i$ and $j$ vanish by symmetry. We are therefore free to order the $\lset$ as an increasing sequence:
\begin{align}\label{eq:MomentWithPermutation}
    \mathtt{r}_{2n}&(L,t)=\frac{(2n)!}{2^{2n}}\sum_{\sigma\in\mathbb{S}_{2n}}\sgn(\sigma)\notag\\
    &\qquad\times \sum_{{\p}_{1}<\cdots<{\p}_{2n}}\sum_{\bm{\xi}_{2n}}\prod_{j=1}^{2n}\delta\left({\p}_{j}-{\p}_{\sigma(j)}+t\xi_{j}\right).
\end{align}
The $\delta$-functions require that $p_{j}-p_{\sigma(j)}=\pm t$, which strongly constrains the permutation $\sigma$. In fact, $\sigma$ is given by a product of 2-cycles, which we prove here. First, suppose that $\sigma$ contains a fixed point, $\sigma(j) = j$. Then~\eqref{eq:MomentWithPermutation} contains a $\delta$-function of the form $\delta(t\xi_{j})$, which cannot be satisfied for $t\neq 0$, hence $\sigma$ cannot contain a fixed point. Next, suppose that $\sigma$ contains a cycle, $\tau$, of length $m>1$. Denote the subset of $\lset$ on which $\tau$ acts by $q_{0}<q_{1}<\cdots<q_{m-1}$. If $\sigma$ contributes to $\mathtt{r}_{2n}$, then the $\delta$-functions require that each $q_n$ can be parameterized as $q_n = q_0 + nt$. Then $\tau$ is constrained to implement
\begin{equation}\label{eq:TauAction}
q_{\tau(n)} = \begin{cases}
    q_{1} & (n=0)\\
    q_{n\pm 1} & (0<n<m)\\
    q_{m-1} & (n=m)\,.
\end{cases}
\end{equation}
However, the $\delta$-functions in~\eqref{eq:MomentWithPermutation} require that $q_{\tau(n)} = q_{n\pm 1}$ for all $n$. This constraint is incompatible with~\eqref{eq:TauAction} unless $m=2$, which implies that $\sigma$ is a product of 2-cycles and we write $\sigma\in\mathbb{S}_{2}^{\otimes n}$. For permutations of this form, $\text{sgn}(\sigma) = (-1)^{n}$ and~\eqref{eq:MomentWithPermutation} can be written as
\begin{align}
    \mathtt{r}_{2n}(L,t)&=(-1)^{n}\frac{(2n)!}{2^{2n}}Q_n(L,t)\label{eq:QDef},
    \\
    Q_n(L,t) &\equiv \sum_{{\p}_{1}<\cdots<{\p}_{2n}}\sum_{\sigma\in\mathbb{S}_{2}^{\otimes n}}\sum_{\bm{\xi}_{2n}}\prod_{j=1}^{2n}\delta(p_{j}-p_{\sigma(j)}+t\xi_{j})\nonumber.
\end{align}
The combinatorial factor $Q_n(L,t)$ then completely determines the SFF,
\begin{align}\label{eq:SFFQExpansion}
    \mK(L,t) = 2^L \sum_{n=0}^{\lfloor L/2\rfloor} \left(\frac{-1}{4}\right)^n Q_n(L,t),
\end{align}
where $\lfloor \cdot\rfloor$ denotes the floor function and $Q_0(L,t) \equiv 1$.

The computation of $Q_{n}(L,t)$ can be interpreted as an embedding problem for dimers. To illustrate this construction, let us begin with the case $t=1$ and consider a one-dimensional chain with $L$ sites labeled $0, 1, \cdots, L-1$. Given a fixed set $\lset$, we say that site $j$ is occupied by a hardcore particle if $j\in\lset$. The $\delta$-functions of Eq.~\eqref{eq:SFFQExpansion} enforce the constraint $p_{j}=p_{\sigma(j)}\pm 1$ for all $j$, that is, each occupied site has at least one occupied neighbor. When this constraint is satisfied, there is a unique choice of $\sigma$ and $\bm{\xi}_{2n}$ which contribute to $Q_{n}$ and we say the particles on sites $j$ and $\sigma(j)$ form a dimer (see Fig.~\ref{fig:cue_pic1} for a diagrammatic representation of contributions to $Q_n$). It follows that $Q_{n}(L,t=1)$ is given by the number of ways to embed $2n$ identical, hardcore particles into $L$ sites such that each particle has at least one neighbor, which is easily counted:
\begin{equation}\label{eq:QAtT1}
    Q_n(L,t=1) = \binom{L-n}{n}.
\end{equation}
Substituting this into Eq.~\eqref{eq:SFFQExpansion}, we obtain the SFF at $t=1$,
\begin{equation}\label{eq:SFFt1}
\begin{aligned}
    \mK(L,1)&=2^L\sum_{n=0}^{\lfloor L/2\rfloor} \left(\frac{-1}{4}\right)^n \binom{L-n}{n} 
    \\
    &= L+1.
\end{aligned}
\end{equation}

\begin{figure}
    \centering
    \includegraphics[width=\columnwidth]{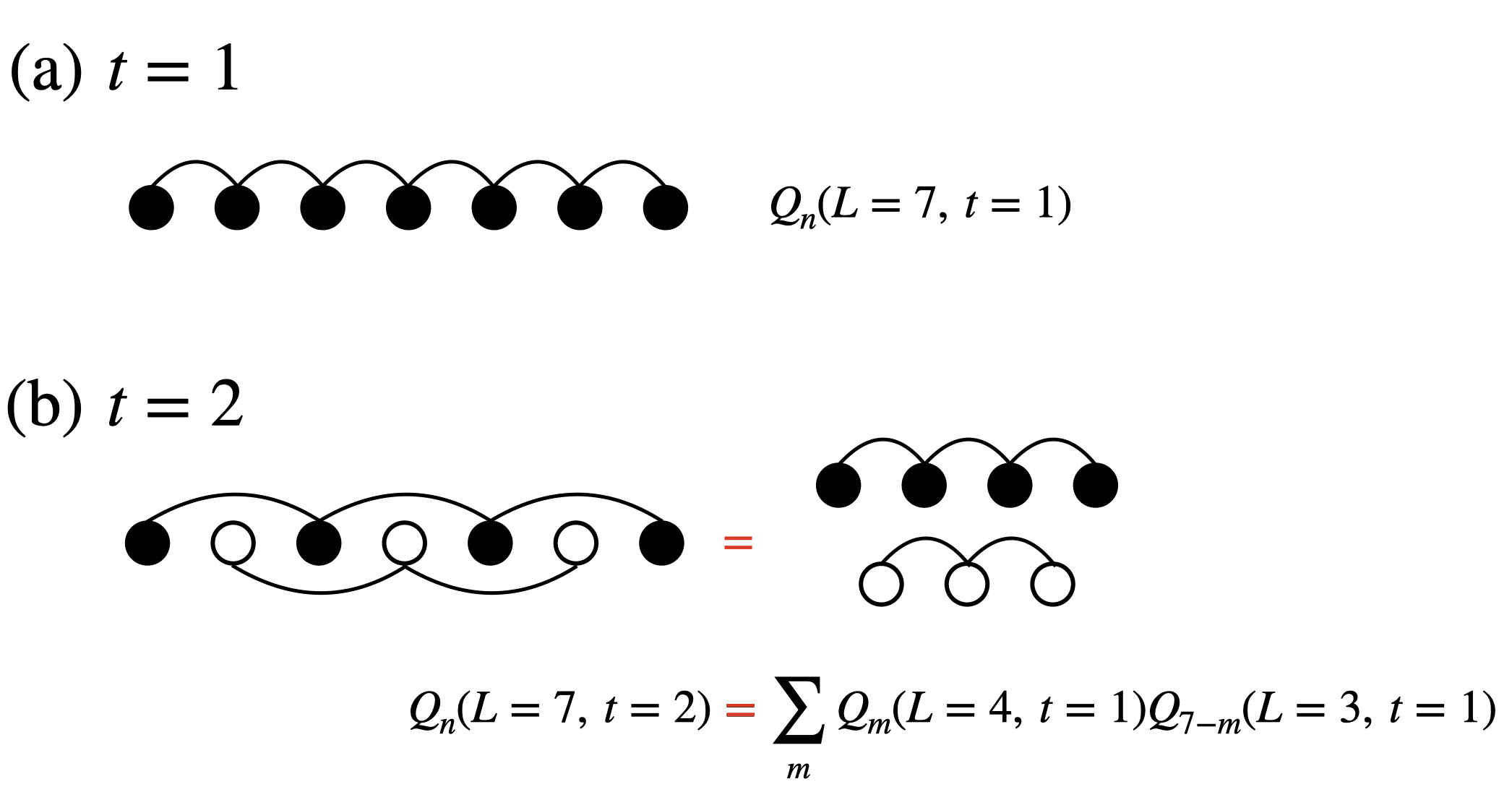}
    \caption{Pictorial representation of dimer embeddings. (a) At $t=1$, all sites are in the same equivalence class and the dimers are built of neighboring particles with counting~\eqref{eq:QAtT1}. (b) For $t=2$, odd and even sites (black and white circles) are decoupled for the dimer problem and the associated counting problem reduces to a correlated pair of $t=1$ problems.}
    \label{fig:cue_pic2}
\end{figure}

This method of determining $Q_n(L,t)$ can be extended to larger values of $t$ in a straightforward manner. The constraints of Eq.~\eqref{eq:QDef} are satisfiable in general when ${\p}_{j}={\p}_{\sigma(j)}\pm t$. Again interpreting the $\lset$ as a list of occupied sites, $Q_n(L,t)$ counts the number of ways to occupy $L$ sites with $2n$ identical hardcore particles such that each particle has at least one neighbor at distance $t$. We refer to a pair of particles at sites $j$ and $\sigma(j) = j\pm t$ in this setting as a range-$t$ dimer, see Fig.~\ref{fig:cue_pic1}(b).

To illustrate the utility of this interpretation, let us consider the case $t=2$ and write $L=2N+r$, where $N$ and $r$ are the quotient and remainder of $L$ with respect to $t$ (see Fig.~\ref{fig:cue_pic2}). In this case, the even and odd sites are decoupled for the purposes of our counting since range-2 dimers are confined to either even or odd sites. By considering them independently, the embedding problem is reduced to a correlated pair of $t=1$ embedding problems. More precisely, we need to sum over all possible ways of embedding $m$ dimers into the even sites and $n-m$ dimers into the odd sites,
\begin{align}
    Q_n(2N+r,2)=\sum_{m=0}^{n}Q_m(N+r,1)Q_{n-m}(N,1)\;.
\end{align}
%
This information together with Eq.~\eqref{eq:QAtT1} is sufficient to compute $\mK(L,t=2)$, although doing so at this stage is unnecessary for our purposes.

The generalization of this argument to arbitrary $t$ is straightforward. First, we decompose $L$ into its quotient and remainder with respect to $t$ by writing $L=Nt+r$. Since sites in different congruence classes modulo $t$ are independent for the dimer problem, this is equivalent to an effective $t=1$ problem with $r$ subsystems of $N+1$ sites and $t-r$ subsystems of $N$ sites. In total, we obtain
\begin{widetext}
\begin{equation}
\begin{aligned}\label{eq:GenericDimerCounting}
    Q_n(L=Nt+r,t)&=\sum_{n_{1},\cdots,n_{r}=0}^{\lfloor (N+1)/2\rfloor}\;\;\sum_{n_{r+1},\cdots,n_{t}=0}^{\lfloor N/2\rfloor}\;\left[\prod_{j=1}^{r} Q_{n_j}(N+1,1) \right]\left[\prod_{k=r+1}^{t} Q_{n_k}(N,1)\right]\delta\left(n-\sum_{l=1}^t n_{l}\right),
\end{aligned}
\end{equation}
\end{widetext}
where $n_1, \dots, n_r$ represent the number of dimers assigned to each of the $(N+1)$-site subsystems and $n_{r+1},\dots,n_t$ to those of the $N$-site subsystems. The $\delta$-function fixes the total number of dimers to equal $n$.
In other words, the computation of $Q_{n}(L,t)$ can always be reduced to a correlated set of $t=1$ embedding problems whose exact solution is a simple modification of Eq.~\eqref{eq:QAtT1}.

At this point, the SFF can be determined by taking the dimer counting~\eqref{eq:GenericDimerCounting}, plugging it into~\eqref{eq:QDef}, and computing the SFF via the moment expansion~\eqref{eq:SFFMomentExpansion}. Rather than carry out these sums explicitly, we will instead use the moment expansion to prove that the SFF satisfies a factorization identity which immediately allows us to compute $\mK(L,t)$ in closed form.

\subsection{Factorization identity for the SFF}\label{sec:Factorization}
In the previous section, we reduced the combinatorial problem of computing $Q_{n}(L,t)$ to a correlated set of exactly solvable combinatorial problems. In this section, we use that result to derive a related factorization identity for the SFF. More precisely, given a system size $L$ and an integer time $0<t\leq L$, decompose $L$ into its quotient and remainder with respect to $t$, $L=Nt+r$. Then the SFF satisfies the factorization identity
\begin{equation}\label{eq:FactorizationIdentity}
\mK(L,t) = \mK(N+1,1)^{r} \mK(N,1)^{t-r}.
\end{equation}
Beginning from the right hand side, equations~\eqref{eq:SFFQExpansion} and~\eqref{eq:GenericDimerCounting} yield
\begin{widetext}
\begin{align}
\mK(N+1,1)^{r} \mK(N,1)^{t-r}
=2^{L}\sum_{n_{1},\cdots, n_{r}=0}^{\lfloor (N+1)/2 \rfloor}\sum_{n_{r+1},\cdots,n_{t}=0}^{\lfloor N/2\rfloor}\left(-\frac{1}{4}\right)^{ \sum_{\ell=1}^t n_\ell  }\prod_{j=1}^{r}Q_{n_j}(N+1,1)\prod_{k=r+1}^{t}Q_{n_k}(N,1).
\end{align}
\end{widetext}
Now we rearrange the sums into terms with a fixed number of dimers $n = \sum_{\ell=1}^t n_\ell$.
To do so, multiply the summand by the constraint $\delta(n-\sum_{\ell=1}^t n_\ell)$ and  sum over all possible values of $n$.
Using Eq.~\eqref{eq:GenericDimerCounting}, we obtain
\begin{equation}
\begin{aligned}
    &\mK(N+1,1)^{r} \mK(N,1)^{t-r}\\
    &= 2^L \sum_{n} \left(\frac{-1}
    {4}\right)^n Q_n(Nt+r,t)
    \\
    &= \mK(Nt+r,t),
\end{aligned}
\end{equation}
where we used Eq.~\eqref{eq:SFFQExpansion} in the second equality.
This completes the proof of the factorization identity~\eqref{eq:FactorizationIdentity}.

\subsection{Closed-form SFF (integer $t$)}\label{ssec:ExactCUE}
Combined with Eq.~\eqref{eq:SFFt1}, the factorization identity~\eqref{eq:FactorizationIdentity} readily gives an exact closed form for the SFF for any $t\in\mathbb{Z}$,
\begin{equation}\label{eq:ExactSFF}
\mK(L,t>0) = (N+1)^t\left( \frac{N+2}{N+1}\right)^{L-Nt}.
\end{equation}
We emphasize that this result was obtained without any approximations and is exact for any $L$ and $t$. To the authors' knowledge, this is the first exact solution for an SFF which exhibits exponential growth~\cite{Liao2020,Winer2020}. With the remainder of this section, we investigate properties of the SFF and connect its exponential growth to random matrix universality in the single-particle sector.

The SFF in Eq.~\eqref{eq:ExactSFF} grows through a sequence of exponential ramps with a growth rate that depends on $L/t$. To see this, fix $L$ and make a list of the divisors of $L$ in ascending order: $1=t_{1}<t_{2}<\cdots <t_{M}=L$. Now choose a time $0<t\leq L$ and find the divisors of $L$ which satisfy $t_{j}\leq t\leq t_{j+1}$. Defining $N_{j} = L/t_{j}\in\mathbb{Z}$, Eq.~\eqref{eq:ExactSFF} yields
\begin{equation}\label{eq:ExactPiecewise}
\mK(L,t) = \mK(L,t_j)\exp\left[\lambda_j(t-t_j)\right]\,,
\end{equation}
where $\lambda_j\equiv (N_j+1)\ln (N_j+1) - N_j\ln(N_j+2)$ is a constant for $t\in\left[t_{j},t_{j+1}\right)$. In particular, the growth rate $\lambda_j$ jumps when $\lfloor L/t\rfloor$ changes. At late times, meaning $L/t\sim O(1)$, most values of $t$ do not divide $L$ and the SFF is simply described in terms of piecewise exponential functions. At early times, $t/L\ll 1$, the density of divisors of $L$ is large and it is convenient to find an alternative representation for the SFF. When $N=L/t\in\mathbb{Z}$,
\begin{equation}\label{eq:K_divt}
\mK(L,t) = \left(\frac{L}{t} +1\right)^{t}\qquad (L/t \in \mathbb{Z}).
\end{equation}
Clearly, the SFF of the single-particle CUE grows exponentially in time, in contrast with the standard CUE~\eqref{eq:SingleParticleSFF2}, which grows linearly in time.

\begin{figure}
    \centering
    \includegraphics[width=\columnwidth]{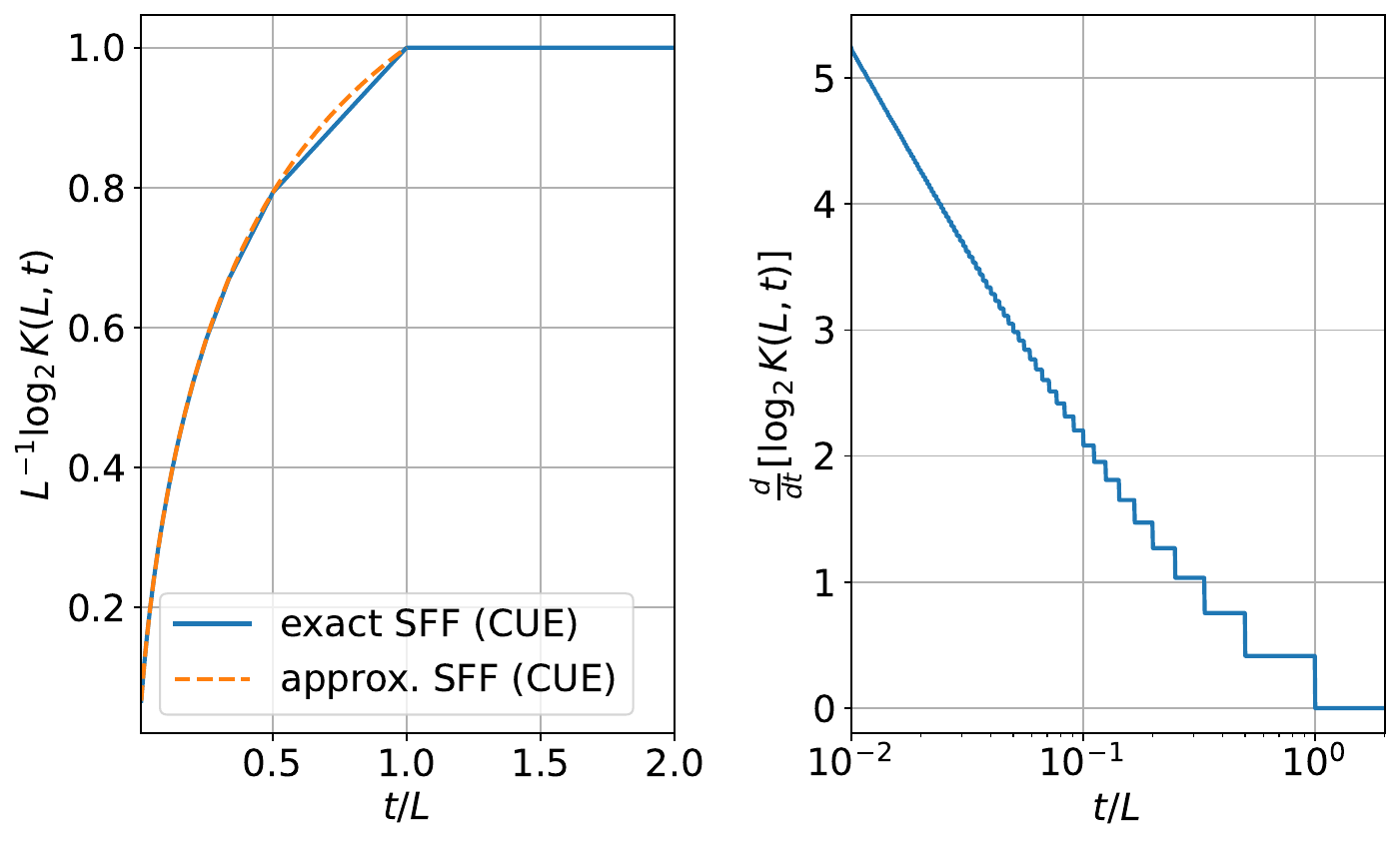}
    \caption{Left: The exact SFF, $\mK(L, t)$ in Eq.~\eqref{eq:ExactSFF} for the CUE (solid lines). The dashed line shows the approximate form of the SFF from Eq.~\eqref{eq:K_divt}. 
    Right: The instantaneous growth rate of the SFF, $\frac{d}{dt} \log_2 \mK(L, t)$. The jumps correspond to the boundaries between exponential ramps, which occur when $t$ divides $L$.}
    \label{fig:exact_cue}
\end{figure}

The exact SFF~\eqref{eq:ExactSFF} can be reorganized to exhibit a scaling collapse. One can easily show that Eq.~\eqref{eq:ExactSFF} can be rewritten as
\begin{equation}\label{eq:K_in_log}
    \frac{\log_2 \mK}{L} = \frac{t}{L}\log_2(N+1)+\left(1-\frac{Nt}{L}\right)\log_2 \left( \frac{N+2}{N+1}\right).
\end{equation}
Here, the RHS of~\eqref{eq:K_in_log} depends only on the ratio $t/L$. As we will see in Sec.~\ref{sec:CSEandCOE}, similar scaling collapses hold for other circular ensembles and serve as an indicator of random matrix statistics at the single-particle level in a many-body system.
Moreover, differentiating Eq.~\eqref{eq:K_in_log} when $t/L\notin\mathbb{Z}$ yields
\begin{align}
    \frac{d}{dt} \log_2 \mK(L, t) = \log_2(N+1) - N \log_2 \left( \frac{N+2}{N+1}\right).
\end{align}
The RHS is piecewise constant and highlights the stepwise exponential growth of the SFF, as illustrated in Fig.~\ref{fig:exact_cue}.

The exponential growth of the many-body SFF can be understood as a consequence of the linear relation between many-body and single-particle quasienergies, $\Theta(\bm{n}) = \bm{n}\cdot\bm{\theta}$, which is absent in interacting systems. This relation implies that dephasing of the single-particle quasienergies is sufficient to destroy all many-body correlations. Given that the average spacing of single-particle quasienergies is given by $2\pi /L$, we expect the SFF~\eqref{eq:SingleParticleSFF2} to saturate at the single-particle Heisenberg time $t_{H}^{(1)}=L$. Nonetheless, the total Hilbert space is still exponentially large in $L$, which implies the exponential growth in Eq.~\eqref{eq:ExactPiecewise}.


\subsection{Exact SFF (non-integer $t$)}\label{ssec:RealtimeMoments}
When $t$ is an arbitrary real number (i.e., not an integer), our evaluation of the moments in Sec.~\ref{sec:MomentsFromDimers} breaks down and the combinatorial interpretation of the moments is lost. Nevertheless, the moment expansion is still analytically exact, although it is gauge-dependent. Throughout this section, we fix the gauge $\theta_j\in\left[-\pi,\pi\right)$, for which the moments~\eqref{eq:MomentSums} are given by
\begin{widetext}
\begin{align}
\mathtt{r}_{n}(L,t)&=\frac{1}{\left(2\pi\right)^n}\sum_{\sigma\in\mathbb{S}_n}\text{sgn}(\sigma)\sum_{{p}_1,\dots,{p}_n=0}^{L-1}\sum_{\bm{\xi}_{n}}\prod_{j=1}^{n}\frac{\sin\left[\pi\left(p_j-p_{\sigma(j)}+t\xi_j\right)\right]}{p_j-p_{\sigma(j)}+t\xi_j}
\\
&=\left(\frac{t\sin\left(\pi t\right)}{\pi}\right)^n\sum_{\sigma\in\mathbb{S}_{n}}\text{sgn}(\sigma)\sum_{{p}_1,\dots,{p}_n=0}^{L-1}\prod_{j=1}^{n}\frac{1}{\left(t-\left(p_{j}-p_{\sigma(j)}\right)\right)\left(t+\left(p_{j}-p_{\sigma(j)}\right)\right)}.\label{eq:GaugedMomentsSum}
\end{align}
\end{widetext}
In the last equality, we have performed the sum over each $\xi_j=\pm 1$. 

Evaluating the moments appears to be non-trivial, but the problem can be simplified by applying symmetries. In particular, the $\left\{p_i\right\}$ sums of~\eqref{eq:GaugedMomentsSum} do not depend on the details of a particular permutation $\sigma$; rather, they are determined entirely by the conjugacy class (or cycle decomposition) of $\sigma$. Each conjugacy class of $\mathbb{S}_{n}$ can contain many permutations, so it is much more efficient to work with the conjugacy classes themselves.

To parameterize the conjugacy classes of $\mathbb{S}_{n}$, we note that there is a one-to-one correspondence between the conjugacy classes and integer partitions of $n$. An integer partition of $n$ is a decomposition of $n$ into a sum of positive integers $\lambda_{i}$ each with multiplicity $m_i$: $n = \sum_{i=1}^{s}m_{i}\lambda_{i}$. We indicate that a tuple forms a partition of $n$ by writing $(\bm{\lambda},\bm{m})\vdash n$.
To make use of this correspondence, we also need the dimension of each conjugacy class, $d(\bm{\lambda},\bm{m})$, which is well-known:
\begin{equation}
    d(\bm{\lambda},\bm{m})=\frac{n!}{\prod_{j}\left[m_{j}!\lambda_{j}^{m_{j}}\right]}
\end{equation}
Using the conjugacy class decomposition, the moments are given by
\begin{equation}
\mathtt{r}_{n}(L,t)=\left(\frac{t\sin\left(\pi t\right)}{\pi}\right)^n\sum_{(\bm{\lambda},\bm{m})\vdash n}\text{sgn}((\bm{\lambda},\bm{m}))d(\bm{\lambda},\bm{m})\prod_{j=1}^{s}q_{\lambda_j}^{m_j}
\end{equation}
where
\begin{equation}\label{eq:Qdef}
q_{\lambda}(L,t) = \sum_{p_{1},\cdots,p_{\lambda}=0}^{L-1}\prod_{j=1}^{\lambda}\frac{1}{\left(t-(p_j-p_{j+1})\right)\left(t+(p_j-p_{j+1})\right)}
\end{equation}
and periodic boundary conditions are assumed: $p_{\lambda+1}\equiv p_{1}$.

\begin{figure}
    \centering
    \includegraphics[width=\columnwidth]{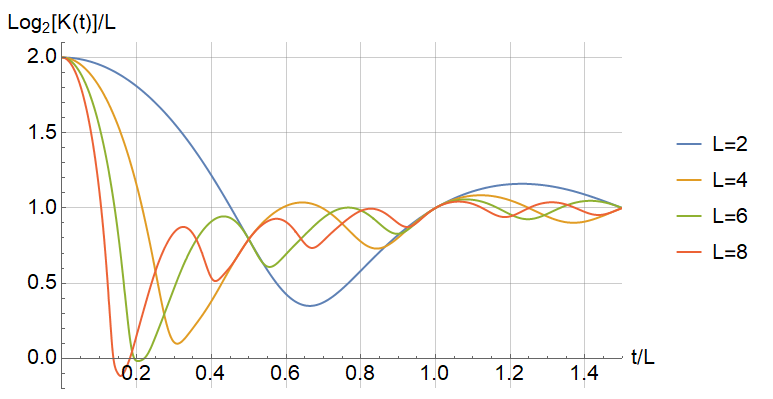}
    \caption{Exact SFF for a range of system sizes and continuous time. We have scaled the axes such that the SFF exhibits a scaling collapse whenever $t$ is an integer. Between the integers, the SFF oscillates and does not have a scaling collapse in general.}
    \label{fig:ContinuousSFF}
\end{figure}

The functions $q_{n}(L,t)$ can be evaluated efficiently using  matrix multiplication. For a fixed choice of $L$, let $T(t)$ be an $L\times L$ matrix with elements
\begin{equation}
\left[T(t)\right]_{jk} = \frac{1}{\left(t-(j-k)\right)\left(t+(j-k)\right)}
\end{equation}
then $q_{n}(L,t) = \tr\left[T(t)^n\right]$, which is much faster to evaluate than the naive sums in~\eqref{eq:Qdef}.

Using this matrix representation, we have evaluated the SFF exactly for a range of system sizes (see Fig.~\ref{fig:ContinuousSFF}). The behavior of the SFF is rich, including an exponential ramp for integer $t$ (see Eq.~\eqref{eq:ExactSFF}) and associated scaling collapse~\eqref{eq:K_in_log}. Between integer arguments, the SFF oscillates and does not exhibit a scaling collapse. Nevertheless, the oscillations are highly constrained since~\eqref{eq:GaugedMomentsSum} implies that the moments can be written as $\mathtt{r}_{n}(L,t) = f_{n}(L,t)\sin(\pi t)^n/g_{n}(L,t)$, where $f_{n}, g_{n}$ are polynomials in $t$.

While the full SFF is difficult to write down in closed form, there are two limits where we can capture its qualitative behavior, namely the limits of short and long times. First, we consider the short time limit $t\ll 1$. In this case, the moments of~\eqref{eq:GaugedMomentsSum} are dominated by terms where $p_j=p_{\sigma(j)}$ for all $j$. The counting of such contributions is straightforward since, via~\eqref{eq:DefMoments},
\begin{equation}
\begin{aligned}
\mathtt{r}_{n}(L,t=0) &= \int d\bm{\theta}_{n}\mathtt{R}_{n}(\bm{\theta}_n)\\
&=\frac{L!}{(L-n)!}.
\end{aligned}
\end{equation}
By counting powers of $t$, we then find
\begin{equation}
\mathtt{r}_{n}(L,t\to 0) = \frac{L!}{(L-n)!}\left(\frac{\sin(\pi t)}{\pi t}\right)^n+\mathcal{O}(t^{4})
\end{equation}
The SFF is then well-approximated at short times by
\begin{equation}
\begin{aligned}
\mK(L,t\to 0) &\approx 2^L\sum_{n=0}^{L}\binom{L}{n}\left(\frac{\sin(\pi t)}{\pi t}\right)^n
\\
&=2^{L}\left[1+\frac{\sin(\pi t)}{\pi t}\right]^L
\end{aligned}
\end{equation}

\begin{figure}
    \centering
    \includegraphics[width=\columnwidth]{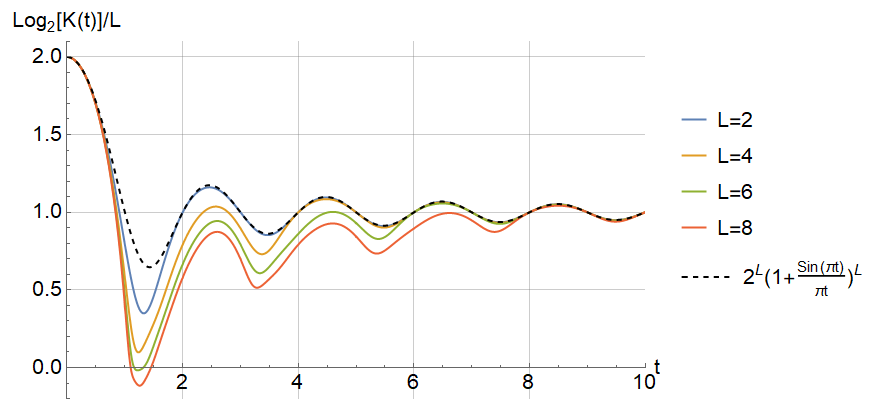}
    \caption{Exact SFF for a range of system sizes and continuous time. We have scaled the axes such that the SFF exhibits scaling collapses at early $(t\ll 1)$ and late $(t\gtrsim L)$ times. In addition both limits are well-approximated by $2^{L}(1+\sin(\pi t)/\pi t)^L$.}
    \label{fig:ContinuousSFFApproximation}
\end{figure}

At late times $t\gtrsim L$, we can approximate the SFF by using the intuition that the $\left\{\theta_j\right\}$ have dephased. We therefore expect that the correlations within the joint probability distribution $P(\bm{\theta})$ can be neglected, so that the SFF simplifies:
\begin{equation}
\begin{aligned}
\mK(L,t\gtrsim L) &\approx 2^L\int\frac{d\bm{\theta}}{\left(2\pi\right)^L}\prod_{j=1}^{L}\left[1+\cos(t\theta_j)\right]
\\
&= 2^L\left[1+\frac{\sin(\pi t)}{\pi t}\right]^L.
\end{aligned}
\end{equation}
Intriguingly, both the short-time and long-time behavior are well-approximated by the same scaling function, as shown in Fig.~\ref{fig:ContinuousSFFApproximation}.

\section{Single-particle COE and CSE}\label{sec:CSEandCOE}
In this section, we analyze the SFF of the single-particle circular symplectic ($\beta=4$) and orthogonal ($\beta=1$) ensembles for integer times. In these cases, we are unable to obtain simple closed-form results for the SFF; however, we will derive moment expansions for the form factors which can be evaluated exactly. These approaches do not require any form of sampling and are therefore significantly more efficient than numerical methods based on exact diagonalization and sampling.

\subsection{Circular symplectic ensemble $(\beta=4)$}\label{sec:CSE}
\subsubsection{Moments and the kernel}\label{sssec:CSEMomentsAndKernel}
First, we discuss the circular symplectic ensemble (CSE) and omit the Dyson index $\beta=4$ throughout this subsection for brevity. The moments~\eqref{eq:DefMoments} are defined in terms of the function $C_{4}$ in Eq.~\eqref{eq:Cfunc}. Expanding in exponentials, $C_{4}$ is given by
\begin{align}\label{eq:C4Expansion}
    &2\cos\left(t\theta_{j}\right)+\cos(t\theta_{j})^{2}\notag\\
    &= e^{it\theta_{j}}+e^{-it\theta_{j}}+\frac{1}{4}\left(2+e^{2it\theta_{j}}+e^{-2it\theta_{j}}\right).
\end{align}
Mirroring our notation for the CUE, we introduce $\xi_{j}=0,\pm 1,\pm 2$.
Then the CSE moments are given by
\begin{equation}
\begin{aligned}\label{eq:MomentsCSE_R}
    \mathtt{r}_{n} 
    &=\int d\bm{\theta}_{n}\;\mathtt{R}_{n}(\bm{\theta}_{n})\sum_{\bm{\xi}_{n}}c(\bm{\xi}_{n})e^{it\bm{\theta}_{n}\cdot\bm{\xi}_{n}}
\end{aligned}
\end{equation}
where $c(\bm{\xi}_{n})$ accounts for the amplitudes associated with each choice of $\xi_{j}$ in~\eqref{eq:C4Expansion}, i.e.,
\begin{align}
    c(\bm{\xi}_n) = \prod_{j=1}^n c(\xi_j),\qquad
    c(\xi) = \begin{cases}
        1/2 & (\xi=0)\\
        1 & (\xi=\pm1)\\
        1/4 & (\xi=\pm2).
    \end{cases}    
\end{align}

To evaluate~\eqref{eq:MomentsCSE_R}, we invoke known results relating the $n$-point function $\mathtt{R}_n(\bm{\theta_n})$ to the CSE kernel~\cite{mehta1991random}. The CSE kernel is given by
\begin{equation}\label{eq:KernelCSE}
    \mathtt{K}_{L4}(\theta) = \frac{1}{2}\begin{pmatrix}
        S_{2L}(\theta) && D_{2L}(\theta)\\
        I_{2L}(\theta) && S_{2L}(\theta)
    \end{pmatrix},
\end{equation}
where the kernel entries are functions defined as
\begin{align}
    S_{2L}(\theta) &\equiv \frac{1}{2\pi}\frac{\sin(L\theta)}{\sin(\theta/2)},\\
    D_{2L}(\theta) &\equiv \frac{d}{d\theta}S_{2L}(\theta),\\
    I_{2L}(\theta) &\equiv \int_{0}^{\theta}d\phi\;S_{2L}(\phi).
\end{align}
We note that $S_{L}(\theta)$ is itself the CUE kernel~\eqref{eq:KernelCUE} and is an even function, while $D_{L}(\theta)$ and $I_{L}(\theta)$ are odd.
Similar to the CUE case, it is convenient to represent each element of the kernel as a Fourier series:
\begin{align}\label{eq:eq:KernelElementsCSE}
    S_{2L}(\theta) &= \frac{1}{2\pi}\sum_{{\p}}e^{i{\p}\theta},\\
    D_{2L}(\theta) &=\label{eq:KernelCSED} \frac{i}{2\pi}\sum_{{\p}}{\p}\;e^{i{\p}\theta},\\
    I_{2L}(\theta) &= \frac{1}{2\pi i}\sum_{{\p}}\frac{e^{ip\theta}}{{\p}}.\label{eq:Idef}
\end{align}
Here and throughout this subsection, the sums over ${\p}$ are taken over the half-integers,
\begin{align}
{\p} = -\frac{2L-1}{2}, -\frac{2L-1}{2}+1,\dots, \frac{2L-1}{2}.
\end{align}
Each of the kernel Fourier series are conveniently represented in the compact notation,
\begin{equation}\label{eq:L4Fourier}
    \mathtt{K}_{L4}^{\alpha\beta}(\theta) = \frac{i^{\beta-\alpha}}{4\pi}\sum_{{\p}}{\p}^{\beta-\alpha} e^{i{\p} \theta},
\end{equation}
where $\alpha,\beta = 0,1$ specify matrix elements of the kernel. 

With the kernel so defined, the $n$-point function $\mathtt{R}_{n}(\bm{\theta}_{n})$ is given by
\begin{equation}
    \mathtt{R}_{n}(\bm{\theta}_{n}) = \qdet{\mathtt{K}_{L4}(\theta_{i}-\theta_{j})}_{i,j = 1,\cdots, n},
\end{equation}
where Qdet denotes the so-called quaternionic determinant introduced by Dyson~\cite{DysonEigenvalueCorrelations} and is defined as follows. Let $\sigma\in\mathbb{S}_n$ be a permutation of $n$ elements and assume that $\sigma$ is a composition of $n_{c}(\sigma)$ (disjoint) cycles, $\sigma = \sigma_{1}\sigma_{2}\cdots\sigma_{n_{c}}$.
Each cycle $\sigma_{j}$ acts on a set of $\lambda_{j}^\sigma$ elements, which we label as $\theta_{j,1},\theta_{j,2},\cdots \theta_{j,\lambda_{j}^\sigma}$. We assume that these elements are ordered with respect to $\sigma_{j}$, meaning that $\sigma_{j}$ enacts the map $\theta_{j,1}\to\theta_{j,2}\to\cdots\to\theta_{j,\lambda_{j}^\sigma}\to\theta_{j,1}$. With these conventions, the quaternionic determinant is given by
\begin{equation}
\begin{aligned}
\mathtt{R}_{n}(\bm{\theta}_{n}) &= \sum_{\sigma\in\mathbb{S}_n}(-1)^{n-n_{c}(\sigma)}
\\
&\times\prod_{j=1}^{n_{c}(\sigma)}\left\{\frac{1}{2}\tr\left[\prod_{k=1}^{\lambda_j^\sigma}\mathtt{K}_{L4}(\theta_{j,k}-\theta_{j,k+1})\right]\right\},
\end{aligned}
\end{equation}
where periodic boundary conditions $\theta_{j,\lambda_{j}^\sigma+1}=\theta_{j,1}$ are assumed. Using~\eqref{eq:MomentsCSE_R}, the moments are given by
\begin{align}\label{eq:MomentClosedForm}
    \mathtt{r}_n &=\sum_{\sigma\in \mathbb{S}_{n}}(-1)^{n-n_{c}(\sigma)}\prod_{j=1}^{n_{c}}\Omega_{\lambda_j^\sigma}(L,t),
\end{align}
where
\begin{equation}
\begin{aligned}\label{eq:DefOmega}
    \Omega_{\lambda}(L,t) &\equiv \frac{1}{2}
    \sum_{\bm{\xi}_{\lambda},\bm{\alpha}_\lambda}c(\bm{\xi}_{\lambda})
    \\
    &\times\int d\bm{\theta}_{\lambda}e^{it\bm{\theta}_{\lambda}\cdot\bm{\xi}_{\lambda}}
    \prod_{k=1}^\lambda \mathtt{K}_{L4}^{\alpha_{k}\alpha_{k+1}}(\theta_{k}-\theta_{k+1}),
\end{aligned}
\end{equation}
where periodic boundary conditions are assumed, $\alpha_{l+1}=\alpha_{1}$, $\theta_{l+1}=\theta_1$, and each $\alpha_k$ is summed from 0 to 1.
The quantity $\Omega_{\lambda}(L,t)$ is determined entirely by the cycle size, $\lambda$, and we refer to $\Omega_{\lambda}$ as the cycle contribution.

The sum over permutations in~\eqref{eq:MomentClosedForm} can be simplified significantly by utilizing symmetries. The summand of~\eqref{eq:MomentClosedForm} depends only on the cycle structure of $\sigma$; as discussed in Sec.~\ref{ssec:RealtimeMoments}, it is much more efficient to carry out sums of this type by summing over the conjugacy classes of $\mathbb{S}_{n}$ rather than the permutations themselves. Using the conjugacy class decomposition, the moments Eq.~\eqref{eq:MomentClosedForm} can be written as
\begin{equation}
\begin{aligned}
    \mathtt{r}_{n} &= \sum_{(\bm{\lambda},\bm{m})\vdash n}(-1)^{n-\sum_{j}m_{j}}d(\bm{\lambda},\bm{m})\prod_{j}\left[\Omega_{\lambda_j}(L,t)^{m_{j}}\right]\\
    &= (-1)^{n}n!\sum_{(\bm{\lambda},\bm{m})\vdash n}\prod_{j}\left[\frac{1}{m_{j}!}\left(\frac{-\Omega_{\lambda_j}(L,t)}{\lambda_{j}}\right)^{m_{j}}\right].
\end{aligned}
\end{equation}
Similarly, the SFF is given by
\begin{align}\label{eq:SFFOmega_partition}
    &\mK (L,t)\notag\\
    &= 2^{2L}\sum_{n=0}^{L}(-1)^{n}\sum_{(\bm{\lambda},\bm{m})\vdash n}\prod_{j}\left[\frac{1}{m_{j}!}\left(\frac{-\Omega_{\lambda_j}(L,t)}{\lambda_{j}}\right)^{m_{j}}\right].
\end{align}
In practice, evaluating Eq.~\eqref{eq:SFFOmega_partition} is significantly simpler than summing over all permutations. The number of integer partitions grows asymptotically as $e^{O(\sqrt{n})}$, whereas the number of permutations scales as $n!\sim n^n$.

\subsubsection{Transfer matrix method for computing $\Omega_\lambda(L,t)$}\label{sec:CSETransfer}

The problem of computing the SFF is now reduced to calculating the cycle contributions~\eqref{eq:DefOmega}. Direct analytic calculations of the cycle contributions seem to be non-trivial; here, we take a different tact and evaluate the cycle contributions with transfer matrix methods. This approach gives a reasonably efficient method for computing the SFF exactly, although it does not provide a closed-form solution for arbitrary $L$.

To put the cycle contributions in a form amenable to transfer matrices, we use the kernel function representation of Eq.~\eqref{eq:L4Fourier} and carry out the $\theta$-integrals, yielding
\begin{widetext}
\begin{align}\label{eq:OmegaSums}
       \Omega_\lambda(L,t\in\mathbb{Z}) &= \frac{1}{2^{\lambda+1}}\sum_{\bm{\xi}_{\lambda}}c(\bm{\xi}_\lambda)\sum_{\{\alpha_{i}=0,1\}}\sum_{{\p}_{1},\dots, {\p}_{\lambda}}{\p}_{1}^{\alpha_{2}-\alpha_{1}}{\p}_{2}^{\alpha_{3}-\alpha_{2}}\cdots {\p}_{\lambda}^{\alpha_{\lambda+1}-\alpha_{\lambda}}\prod_{j=1}^{\lambda}\delta({\p}_{j}-{\p}_{j-1}+t\xi_{j}),
\end{align}
\end{widetext}
where periodic boundary conditions are assumed throughout and we have used the fact that $t\in\mathbb{Z}$.

The representation of the cycle contribution in Eq.~\eqref{eq:OmegaSums} can be thought of as a sum of weighted one-dimensional walks consisting of $\lambda$ steps. In this interpretation, we regard the index $p_{n}$ as the coordinate of a walker after $n-1$ steps. The $\delta$-functions enforce the constraint that the walker takes steps of size $t\xi_{j}$, where each choice of $|\xi_{j}|$ comes with a weight encoded by the function $c(\xi_{j})$. The $\delta$-functions also enforce the constraint $t\sum_{i}\xi_{i}=0$, meaning that each walk terminates on the site where it began. The cycle contribution is then given by a sum of weights encoded by the steps $\xi_{j}$ and coordinates $p_{n}$.

The random walk picture motivates us to define a $2L\times 2L$ transfer matrix $T$,
\begin{align}
    T_{{\p} {\p}'} \equiv \sum_\xi  c(\xi) \delta({\p} - {\p}' +t \xi),
\end{align}
whose elements are $1/2$ on the diagonal, $1$ on diagonals $t$ and $-t$, and $1/4$ on diagonals $2t$ and $-2t$. With this representation, we have, for example,
\begin{equation}
\sum_{\bm{\xi}_{2}}c(\bm{\xi}_{2})\sum_{p_{1},p_{2}}\delta(p_{1}-p_{2}+t\xi_{1})\delta(p_{2}-p_{1}+t\xi_{2}) = \tr\left(T^{2}\right)
\end{equation}
In other words, the transfer matrix $T$ evolves the walker from one site to another, keeping track of the weight which comes from the steps $\xi_{j}$. Computing the trace of $T^{m}$ then amounts to summing those weights for all length-$m$ random walks which begin and end on the same site.

To incorporate the effect of the coordinates $p_n$ on the weights, we regard the $\alpha$-indices as internal degrees of freedom of the walker. Then it is natural to define an enlarged $4L\times 4L$ transfer matrix, $\mT$, with elements
\begin{align}\label{eq:bigT}
    \mT_{\alpha {\p}, \alpha' {\p}'} = {\p}^{\alpha'-\alpha} T_{{\p} {\p}'}.
\end{align}
In terms of $\mT$, the cycle contribution simplifies to
\begin{align}\label{eq:OmegaTransfer}
    \Omega_\lambda(L,t) = \frac{1}{2^{\lambda+1}} \tr\left(\mT^{\lambda}\right).
\end{align}
While the transfer matrix formalism might seem like a simple rewriting of the cycle contribution, in practice it is exponentially faster to compute Eq.~\eqref{eq:OmegaTransfer} rather than the naive sums of Eq.~\eqref{eq:OmegaSums}. This is analogous to computational speedups achieved with matrix product states more generally~\cite{Orus_TN,Schollwoeck_DMRG}.

We conclude this section by estimating the resources necessary to evaluate the full SFF. First, since $\mK(L,t\geq 2L) = 6^{L}$ (see Sec.~\ref{ssec:SingleParticleCSEHeisenbergTime}), it is sufficient to compute $\mK(L,t)$ for $t=1,2,\cdots, L-1$. Computing a cycle contribution $\Omega_{\lambda}(L,t)\propto\tr\left(\mT^{\lambda}\right)$ to the $n$th moment takes $O(L^2\lambda)\lesssim O(L^2 n)$ operations. To compute the $n$th moment, we need to sum over all $O(e^{\sqrt{n}})$ integer partitions of $n$, which have an average length that scales as $O(\sqrt{n}\log n)$~\cite{PartitionLengths}. The operations needed to compute $\mathtt{r}_{n}$ then scale as $O(L^{2}n^{3/2}e^{\sqrt{n}}\log n)$. Computing the SFF for a single value of $t$ then takes approximately
\begin{equation}\label{eq:CSEComplexity}
O\left(L^{2}\int_{0}^{L}dx\; e^{\sqrt{x}}x^{3/2}\log x\right)\sim O\left(L^{4}e^{\sqrt{L}}\log L\right)
\end{equation}
operations. Computing the SFF for $t=1,2,\cdots,L$ then gives the overall scaling $O(L^{5}e^{\sqrt{L}}\log L)$.

\subsubsection{Numerically exact SFF}\label{sec:ExactCSE}

\begin{figure}
    \centering
    \includegraphics[width=\columnwidth]{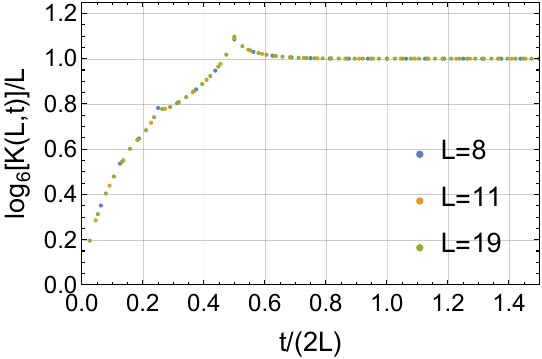}
    \caption{Exact SFF for the CSE calculated numerically with the random walk picture.}
    \label{fig:exact_cse}
\end{figure}

In this section, we present numerically exact results for the SFF of the single-particle CSE obtained using the transfer matrix method developed in Sec.~\ref{sec:CSETransfer}. Figure~\ref{fig:exact_cse} shows the SFF, $\mK(L, t)$, for system sizes $L = 8, 11, 19$, plotted as a function of the scaled time $t/L$. The SFF is normalized by a factor of $L$ and plotted on a logarithmic scale. Here we use $\log_6$ to emphasize the late-time behavior $\mK (L,t\ge 2L)=6^L$.

Several key features of the SFF are apparent in Fig.~\ref{fig:exact_cse}. First, for all system sizes, the SFF exhibits a clear exponential ramp at early times, $t \ll L$. This is consistent with the exponential ramp found analytically for the CUE in Sec.~\ref{ssec:ExactCUE} and highlights that exponential growth of the SFF is a generic feature of the single-particle circular ensembles. Second, the SFF saturates to a plateau value of 1 at late times, $t \gtrsim L$. The plateau indicates that the CSE has the single-particle Heisenberg time, $t_H^{1,\mathrm{(CSE)}} =2L$, which we show analytically in Sec.~\ref{ssec:SingleParticleCSEHeisenbergTime}.

Comparing the curves for different $L$, it is clear that they exhibit a scaling collapse when plotted against the scaled time $t/L$. This suggests that, for the system sizes accessible to our transfer matrix method, the SFF of the CSE depends on $L$ and $t$ exclusively through their ratio. An analogous scaling collapse was proven exactly for the single-particle CUE in Sec.~\ref{ssec:ExactCUE}.

The numerical results of Fig.~\ref{fig:exact_cse} definitively establish that the single-particle CSE exhibits an exponential ramp in the SFF. Together with the exact CUE solution and analogous COE numerics discussed below in Sec.~\ref{sec:COE}, our results suggest that the exponential ramp is a universal feature of the single-particle circular ensembles and is a signature of random matrix statistics at the single-particle level.

\subsubsection{Single-particle Heisenberg time}\label{ssec:SingleParticleCSEHeisenbergTime}

To illustrate our formalism in practice, we will now prove that the SFF exhibits the late-time behavior $\mK(L,t\geq 2L) = 6^{L}$ (assuming $t\in\mathbb{Z})$. This result puts a lower bound on the single-particle Heisenberg time, $t_H^{1,(\text{CSE})}\geq 2L$; in fact, numerics show that this bound is an equality.

First, we note that the cycle contributions simplify at late times. Beginning from Eq.~\eqref{eq:OmegaSums}, we find that $\Omega_{\lambda}(L,t\geq 2L) = L/2^{\lambda}$. The SFF of~\eqref{eq:SFFOmega_partition} then becomes
\begin{align}
    &\mK (L,t\geq2L)\notag\\
    &= 2^{2L}\sum_{n=0}^{L}(-1)^{n}\left(\frac{1}{2}\right)^n\sum_{(\bm{\lambda},\bm{m})\vdash n}\prod_{j}\left[\frac{1}{m_{j}!}\left(\frac{-L}{\lambda_{j}}\right)^{m_{j}}\right].\label{eq:K_CSE_larget_t}
\end{align}
This result seems complicated at first glance, but it can be simplified using techniques employed to solve combinatorial problems involving integer partitions. Inspired by these methods, we will now prove the following identity:
\begin{align}\label{eq:moments_eq}
    \sum_{(\bm{\lambda},\bm{m})\vdash n}\prod_{j}\left[\frac{1}{m_{j}!}\left(\frac{-L}{\lambda_{j}}\right)^{m_{j}}\right]
    = (-1)^n \binom{L}{n}
\end{align}
for all $n$.
\begin{proof}
We make use of the identity
\begin{align}\label{eq:binom_eq}
    (-1)^n \binom{L}{n}
    =\sum_{l=1}^n\frac{(-L)^l}{l!}\sum_{k_1+\dots+ k_l=n}\frac{1}{k_1\cdots k_l},
\end{align}
where each $k_j$ is a positive integer.
The $k_j$ are not necessarily distinct and is is convenient to organize them into ordered pairs $(\lambda_{j},m_{j})$, where each $\lambda_j$ is unique and $m_j$ is its multiplicity.
By definition, we have
\begin{align}
    n = \sum_j \lambda_j m_j;\quad
    \frac{1}{k_1\cdots k_l}
    = \frac{1}{\prod_{j} (\lambda_j)^{m_j}},
\end{align}
which simplifies~\eqref{eq:moments_eq} to
\begin{align}\label{eq:binom_eq}
    (-1)^n \binom{L}{n}
    =\sum_{l=1}^n\frac{(-L)^l}{l!}\sum_{\substack{(\bm{\lambda},\bm{m})\vdash n\\ (\sum_j m_j =l) }}\mathcal{N}(\bm{m})
    \frac{1}{\prod_{j} (\lambda_j)^{m_j}}.
\end{align}
Here we have used the fact that
\begin{align}\label{eq:sum_reprlacement}
\sum_{k_1+\dots+ k_l=n}
    =
\sum_{\substack{(\bm{\lambda},\bm{m})\vdash n\\ (\sum_j m_j =l) }}\mathcal{N}(\bm{m}),
\end{align}
where
\begin{align}\label{eq:N_combinatorics}
    \mathcal{N}(\bm{m})=\frac{l!}{\prod_j m_j!}
\end{align}
accounts for the multiplicity of terms which appear in the sums over each $k_j$.
Substituting Eq.~\eqref{eq:N_combinatorics} into Eq.~\eqref{eq:binom_eq} and gathering the two sums, we have
\begin{align}
    (-1)^n \binom{L}{n}
    &=\sum_{l=1}^n \sum_{\substack{(\bm{\lambda},\bm{m})\vdash n\\ (\sum_j m_j =l) }}
    \prod_{j}\left[ \frac{1}{m_j!} \left(\frac{-L}{\lambda_j}\right)^{m_j}\right]\\
    &= \sum_{(\bm{\lambda},\bm{m})\vdash n}
    \prod_{j}\left[ \frac{1}{m_j!} \left(\frac{-L}{\lambda_j}\right)^{m_j}\right].
\end{align}
This completes the proof of~\eqref{eq:moments_eq}.
\end{proof}
Finally,
combining Eqs.~\eqref{eq:moments_eq} and \eqref{eq:K_CSE_larget_t} and using the binomial theorem, we obtain
\begin{align}
    \mK (L,t\geq2L)
    &= 2^{2L}\sum_{n=0}^{L}\left(\frac{1}{2}\right)^n\binom{L}{n}= 6^L.
\end{align}
This result puts a lower bound on the single-particle Heisenberg time of $2L$. Our numerical results indicate that this bound is tight.

\subsection{Circular orthogonal ensemble $(\beta=1)$}\label{sec:COE}
\subsubsection{Moments and the kernel}

Here, we discuss the circular orthogonal ensemble (COE) and omit the Dyson index $\beta=1$ throughout this subsection. Using the $C$-function in Eq.~\eqref{eq:Cfunc}, the moments of Eq.~\eqref{eq:DefMoments} are given by
\begin{align}
    \mathtt{r}_{n}(L,t) &\equiv     \frac{1}{2^{n}}\sum_{\bm{\xi}_{n}}\int d\bm{\theta}_{n}\; \mathtt{R}_{n}(\bm{\theta}_{n})\;e^{it\bm{\theta}_{n}\cdot\bm{\xi}_{n}},\label{eq:COEMoments}
\end{align}
where $\xi_i=\pm1$.

To evaluate the moments, we again invoke a kernel representation for the $n$-point function $\mathtt{R}_n(\bm{\theta}_n)$.
The COE kernel has the matrix representation~\cite{mehta1991random}
\begin{equation}\label{eq:L1def}
    \mathtt{K}_{L1}(\theta) = \begin{pmatrix}
        S_{L}(\theta) && D_{L}(\theta)\\
        J_{L}(\theta) && S_{L}(\theta)
    \end{pmatrix},
\end{equation}
where $S_L$ and $D_L$ were defined in~\eqref{eq:eq:KernelElementsCSE}, \eqref{eq:KernelCSED} and $J$ is given by
\begin{equation}
    J_{L}(\theta) = -\frac{1}{2\pi i}\sum_{\q}\frac{e^{i\q\theta}}{\q},
\end{equation}
where $\q=\pm(L+1)/2, \pm(L+3)/2, \cdots$, runs over an infinite range.
Recall that the Fourier series of $S_L(\theta)$ and $D_L(\theta)$ have a finite number of terms summed over the range $\p=-\frac{L-1}{2},-\frac{L+1}{2},\dots,\frac{L-1}{2}$. To avoid writing out these ranges explicitly, it is convenient to develop a unified notation for the kernel matrix elements,
\begin{align}\label{eq:L1Fourier}
    \mathtt{K}^{\alpha\beta}_{L1}(\theta) = \frac{i^{\beta-\alpha}}{2\pi}\sum_{\p\in\mathbb{Z}+1/2}\eta_{\alpha\beta}(\p)\p^{\beta-\alpha}e^{i\p \theta},
\end{align}
where $p$ is summed over the half-integers, $\alpha,\beta = 0,1$ index matrix elements of the kernel and
\begin{align}
    \eta_{\alpha\beta}(\p)&= \begin{cases}
        \Theta\left( |\p| - \frac{L}{2} \right) & \alpha=1\ \text{and}\ \beta=0 \\
        \Theta\left( \frac{L}{2} - |\p| \right) & \text{otherwise}
    \end{cases}
\end{align}
where $\Theta(x)$ is the Heaviside function.

The arguments of Sec.~\ref{sssec:CSEMomentsAndKernel} also apply to the COE, yielding an identical result for the SFF in terms of the cycle contributions,
\begin{align}\label{eq:SFFOmega_partition_COE}
    &\mK (L,t)\notag\\
    &= 2^{L}\sum_{n=0}^{L}(-1)^{n}\sum_{(\bm{\lambda},\bm{m})\vdash n}\prod_{j}\left[\frac{1}{m_{j}!}\left(\frac{-\Omega_{\lambda_j}(L,t)}{\lambda_{j}}\right)^{m_{j}}\right].
\end{align}
The difference between the COE and CSE appears in the cycle contributions themselves,
\begin{align}\label{eq:DefOmegaCOE}
    \Omega_{\lambda}(L,t) &\equiv \frac{1}{2^{\lambda+1}}
    \sum_{\bm{\xi}_{\lambda},\bm{\alpha}_\lambda} \int d\bm{\theta}_{\lambda}e^{it\bm{\theta}_{\lambda}\cdot\bm{\xi}_{\lambda}}
    \prod_{k=1}^\lambda \mathtt{K}_{L1}^{\alpha_{k}\alpha_{k+1}}(\theta_{k}-\theta_{k+1}).
\end{align}
Compared to the corresponding CSE expression~\eqref{eq:DefOmega}, the COE cycle contributions are simplified by the fact that the $C$-function of Eq.~\eqref{eq:Cfunc} takes only one value, leading to the prefactor $1/2^{\lambda+1}$.
Nevertheless, the COE is more complex due to the unbounded sums associated with $J_L(\theta)$.

\subsubsection{Transfer matrix method for computing $\Omega_\lambda(L,t)$}\label{sec:COETransfer}
Here we compute the cycle contribution $\Omega_\lambda(L,t)$ by using transfer matrix methods. Following our approch for the CSE, we plug the Fourier series representation of the kernel~\eqref{eq:L1Fourier} into the cycle contributions~\eqref{eq:DefOmegaCOE} of and perform the $\theta$-integrals (assuming $t\in\mathbb{Z}$), obtaining
\begin{align}\label{eq:Omega_and_bigT}
    \Omega_\lambda(L,t)=\frac{1}{2^{\lambda+1}}\tr(\mT^\lambda),
\end{align}
where the transfer matrix $\mT$ has matrix elements
\begin{align}\label{eq:bigT_COE}
    \mT_{\alpha \p, \alpha' \p'} \equiv (-1)^{\alpha(1-\alpha')}\eta_{\alpha \alpha'}(\p)\p^{\alpha'-\alpha} \sum_{\xi=\pm1} {\delta}(\p-\p' + t\xi).
\end{align}
Again, we interpret $\alpha$ as an internal degree of freedom of a walker and $p, p'$ as the walker's coordinates. Then the transfer matrix evolves the walker through steps of size $\pm t$ with appropriate weights such that~\eqref{eq:Omega_and_bigT} holds.

Formally, $\mT$ is infinite-dimensional due to the unbounded sums associated with $J_{L}(\theta)$. However, we will show that the cycle contribution can be computed exactly by truncating $\mT$ such that
\begin{align}\label{eq:Ncut}
    |\p| \le \frac{L-1}{2} + t,
\end{align}
which enables the use of standard methods for any finite $L$ and integer $t$.

The truncation of $\mT$ is justified by analyzing the set of random walks which contribute to~\eqref{eq:Omega_and_bigT}. First note $\Omega_{1}(L,t)=0$, so take $\lambda\geq 2$. Then Eq.~\eqref{eq:Omega_and_bigT} can be written as 
\begin{align}\label{eq:Omega_walks}
    \Omega_\lambda(L,t) = \frac{1}{2^{\lambda+1}}\sum_{\alpha_1,\p_1}\dots \sum_{\alpha_\lambda,\p_\lambda}\prod_{k=1}^\lambda \mT_{\alpha_k p_k,\alpha_{k+1}p_{k+1}},
\end{align}
where $\alpha_{\lambda+1}=\alpha_1$ and $\p_{\lambda+1}=\p_1$. The walks which contribute to~\eqref{eq:Omega_walks} begin and end on the same site. Now suppose there is a $p_{k}$ such that $|p_k|\geq (L+1)/2$. Then for the walk to give a nonzero contribution, $(\alpha_k,\alpha_{k+1})=(1,0)$ in order to satisfy $\eta_{\alpha_k \alpha_{k+1}}(\p_k)\neq0$. Further, since $\alpha_{k+1}=0$, we require that $|p_{k+1}|\le (L-1)/2$. Since $p_{k+1} = p_{k}\pm t$, it follows that $|p_{k}|\leq (L-1)/2 + t$.

In summary, the cycle contribution $\Omega_\lambda(L,t)$ can be obtained from Eq.~\eqref{eq:Omega_and_bigT} and truncating the transfer matrix with the condition~\eqref{eq:Ncut}. This is the main result of this section and, when combined with Eq.~\eqref{eq:SFFOmega_partition_COE}, allows us to obtain the exact SFF numerically. 

The computational cost of computing the SFF can be estimated by the same analysis that led to Eq.~\eqref{eq:CSEComplexity}, with the modification that $\text{dim}(\mT) = L+2t$ is time-dependent. This effect is easily incorporated and we find that computing the SFF at a single value of $t$ takes time $O((L+2t)^2 L^2 e^{\sqrt{L}}\log L)$. For times $t\leq L$, the complexity is dominated by the number of integer partitions of $L$. For $t>L$ the exact calculation continues to become more difficult, but we will show how to obtain late time asymptotic behavior which provides an accurate approximation to the SFF in this case.

\subsubsection{Numerically exact SFF}\label{sec:ExactCOE}
\begin{figure}
    \centering
    \includegraphics[width=\columnwidth]{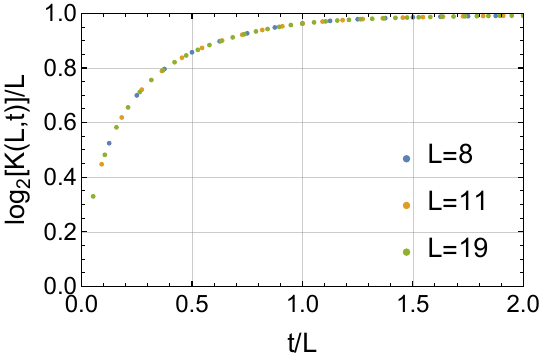}
    \includegraphics[width=0.9\columnwidth]{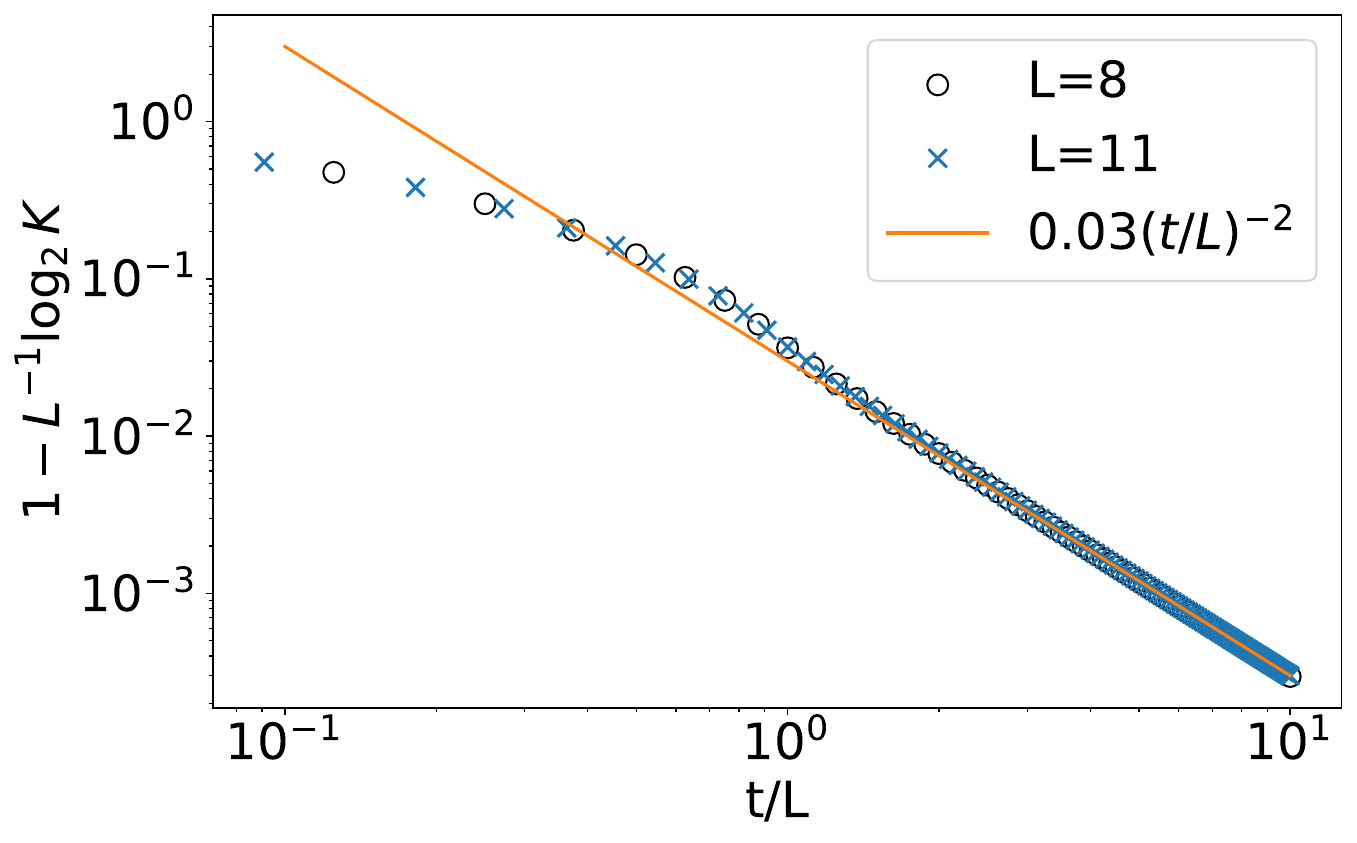}
    \caption{(Top) Exact SFF for the COE calculated numerically with the transfer matrix method. (Bottom) Deviation of the exact SFF $L^{-1}\log_2 \mK(L,t)$ from its asymptotic value 1 for $L=8$ (circle) and $L=11$ (cross). The solid line shows a long-time fit $0.03\times (t/L)^{-2}$.}
    \label{fig:exact_coe}
\end{figure}

In this section, we present numerically exact results for the SFF of the single-particle COE obtained using the transfer matrix method developed in Sec.~\ref{sec:COETransfer}. Figure~\ref{fig:exact_coe} shows the SFF, $\mK(L, t)$, for system sizes $L = 8, 11, 19$, plotted as a function of the scaled time $t/L$. The SFF is normalized by a factor of $L$ and plotted on a logarithmic scale.

The SFF of the COE exhibits several regimes. First, at early times $t \ll L$, the SFF grows more quickly than the exponential ramp found in the CUE. In fact, the growth rate seems to diverge as $t/L\to 0$, although our numerical results do not allow for a definitive statement about the early-time behavior. Second, at intermediate times $t \sim L$, there is a clear crossover to a slower growth rate. The SFF eventually relaxes towards the plateau, although there is no sharp crossover to this regime.

Similar to the CSE, the SFF curves for different system sizes exhibit a clear scaling collapse when plotted against the scaled time $t/L$. This suggests that the SFF of the COE depends on $L$ and $t$ primarily through their ratio. 
Finding an analytical form of the universal scaling collapse function in the $L\to\infty$ limit is an interesting direction for future work.

Finally, the long-time asymptotic behavior of the SFF is qualitatively different from the CUE and CSE cases. In the latter two cases, the SFF becomes constant for integer $t>L_{\beta}$. In contrast, the SFF does not reach saturate for any finite $t$. Based on numerics, we have identified the long-time asymptotic behavior to be $1-L^{-1}\log_2 \mK \approx C \left(\frac{t}{L}\right)^2$, and the coefficient $C$ is estimated as $C\approx 0.030$ by fitting (see Fig.~\ref{fig:exact_coe}). This scaling is equivalent to $\mK(L,t)\approx 2^{L[1-C(t/L)^{-2}]}$.

In summary, our numerically exact results establish that the SFF of the single-particle COE exhibits a novel form of intermediate spectral statistics. The SFF is consistent with random matrix universality at the single-particle level, but shows a nontrivial crossover behavior at early-to-intermediate times. Our results demonstrate the rich variety of spectral statistics possible in single-particle chaotic systems depending on their symmetry structure.

\section{Quantum circuit realizations}\label{sec:circuits}
In this section, we present ensembles of quantum circuits whose form factors converge to those of the single-particle circular ensembles. To simulate fermions in a quantum circuit, we employ nearest-neighbor matchgates, which can be mapped onto non-interacting fermions through the Jordan-Wigner transformation ~\cite{Jozsa2008,Brod2016}. A 2-qubit matchgate has matrix representation
\begin{equation}
G(A,B) = \begin{pmatrix}
A_{11} && 0 && 0 && A_{12}\\
0 && B_{11} && B_{12} && 0\\
0 && B_{21} && B_{22} && 0\\
A_{21} && 0 && 0 && A_{22}
\end{pmatrix}
\end{equation}
and satisfies $\det A = \det B$. For simplicity, we consider  U$(1)$-conserving matchgates, whose matrix representations are of the form
\begin{align}\label{eq:Vmat}
    V_{i,i+1}=\begin{pmatrix}
e^{i\phi_{0}} & 0 & 0 & 0\\
0 & u_{11} & u_{12} & 0\\
0 & u_{21} & u_{22} & 0 \\
0 & 0 & 0 & e^{i \phi_3}
\end{pmatrix}
=\includegraphics[width=1.5cm,valign=c]{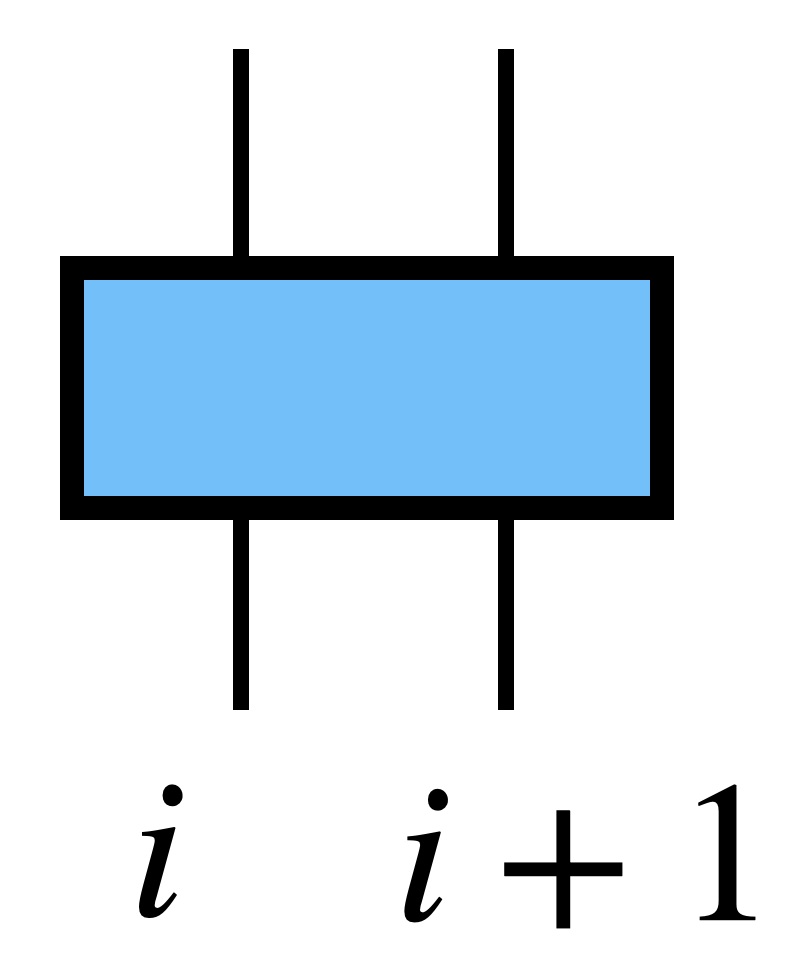}
\end{align}
with the constraint $\mathrm{det}(u)=e^{i(\phi_0+\phi_3)}$. Through the Jordan-Wigner transformation, such a unitary operator is mapped to
\begin{align}
V_{i,i+1}=\exp\left[-i \begin{pmatrix} c_i^\dag & c_{i+1}^\dag \end{pmatrix}
\hat{h}
\begin{pmatrix} c_i \\ c_{i+1}\end{pmatrix}
\right],
\end{align}
where $c_i, c_i^\dag$ are spinless fermionic operators on site $i$ and $\hat{h}$ is a $2\times 2$ Hermitian matrix uniquely determined (modulo gauge transformations) by $V_{i,i+1}$ in Eq.~\eqref{eq:Vmat}~\cite{Brod2016}. A circuit constructed out of matchgates acting on $L_{\beta}$ qubits can be written as
\begin{align}\label{eq:U_and_H}
    \mU=e^{-i \vec{c}^\dag \cdot H\cdot \vec{c}},
\end{align}
where $\vec{c} = (c_1,c_2,\dots,c_{L_\beta})^T$ and $H$ is an ${L_\beta}\times {L_\beta}$ Hermitian matrix that we interpret as a Hamiltonian. In the eigenbasis of $H$, $H \vec{\psi}^{(j)} = \epsilon_j \vec{\psi}^{(j)}$ $(j=1,2,\dots,{L_\beta})$ and Eq.~\eqref{eq:U_and_H} can be written as
\begin{align}\label{eq:U_matchgates}
    \mU = \exp\left[-i\sum_{j=1}^{L_\beta} \epsilon_j n_j\right],
\end{align}
where $n_j=d_j^\dag d_j$, where $d_j = \sum_{k=1}^{L_\beta} (\psi^{(j)}_k)^* c_k$. The circuit unitary~\eqref{eq:U_matchgates} takes the same form as the many-body unitaries of Eq.~\eqref{eq:ManyBodyFloquet}, although the $\epsilon_j$ are determined by the matchgate parameters and circuit geometry.

With these considerations, we conjecture that there exist ensembles of matchgate circuits whose form factors are given by those of the single-particle circular ensembles. Here, we construct such ensembles by using random matchgates with symmetry properties appropriate to each circular ensemble and numerically demonstrate their agreement with the results of Secs.~\ref{sec:CUE},~\ref{sec:CSEandCOE}.

\begin{figure*}
    \centering
    \includegraphics[width=0.66\columnwidth]{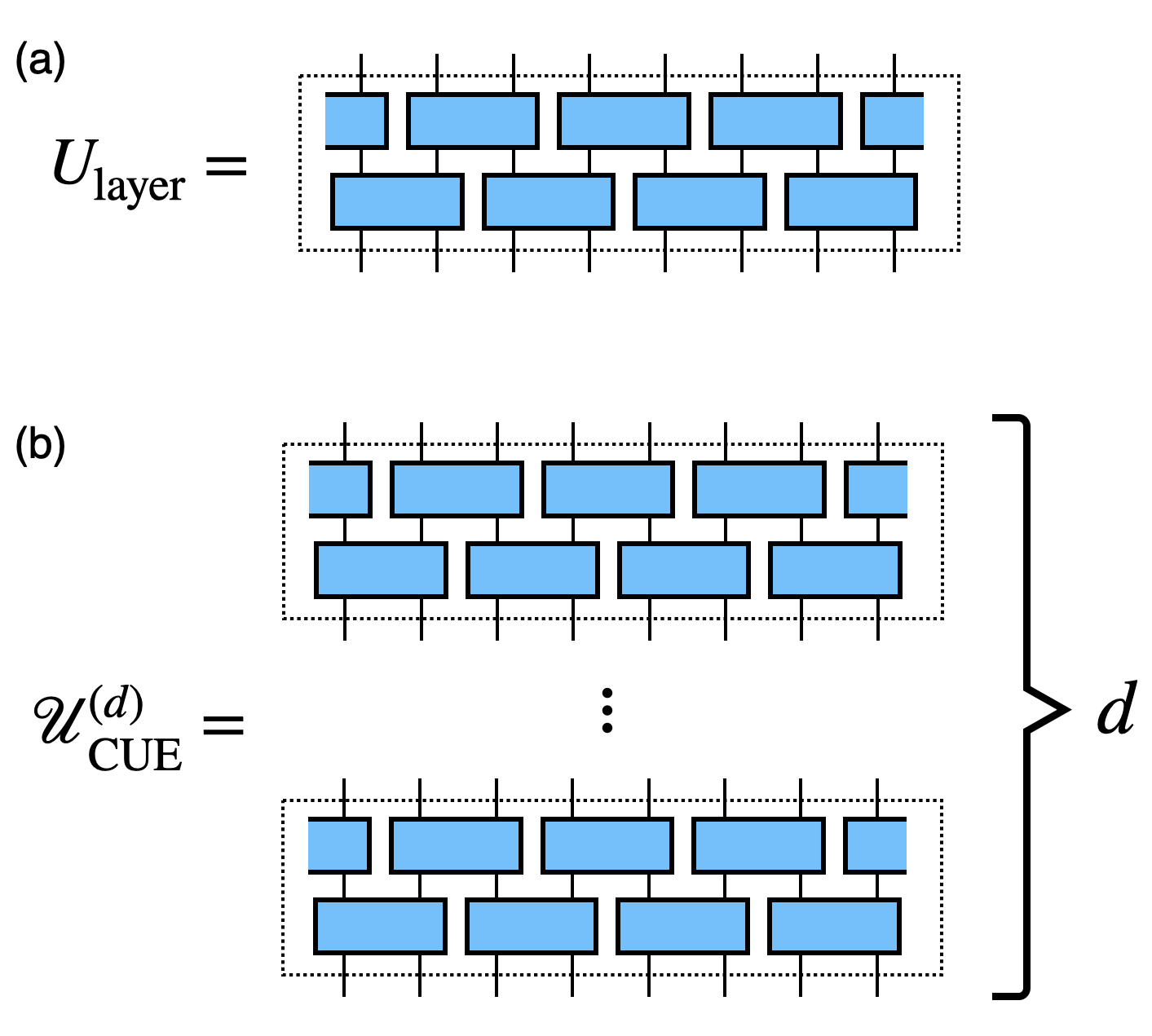}
        \includegraphics[width=0.66\columnwidth]{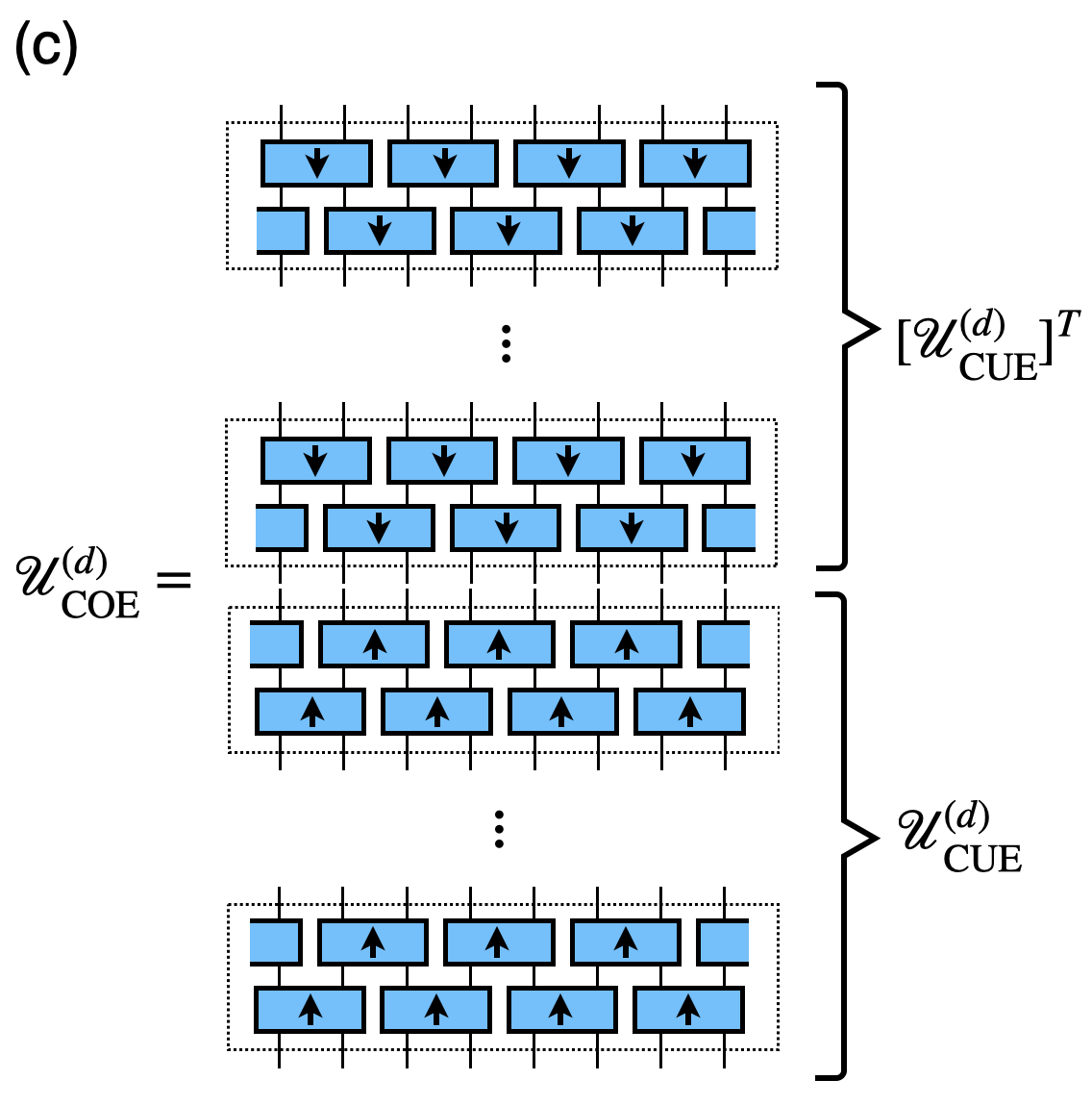}
    \includegraphics[width=0.66\columnwidth]{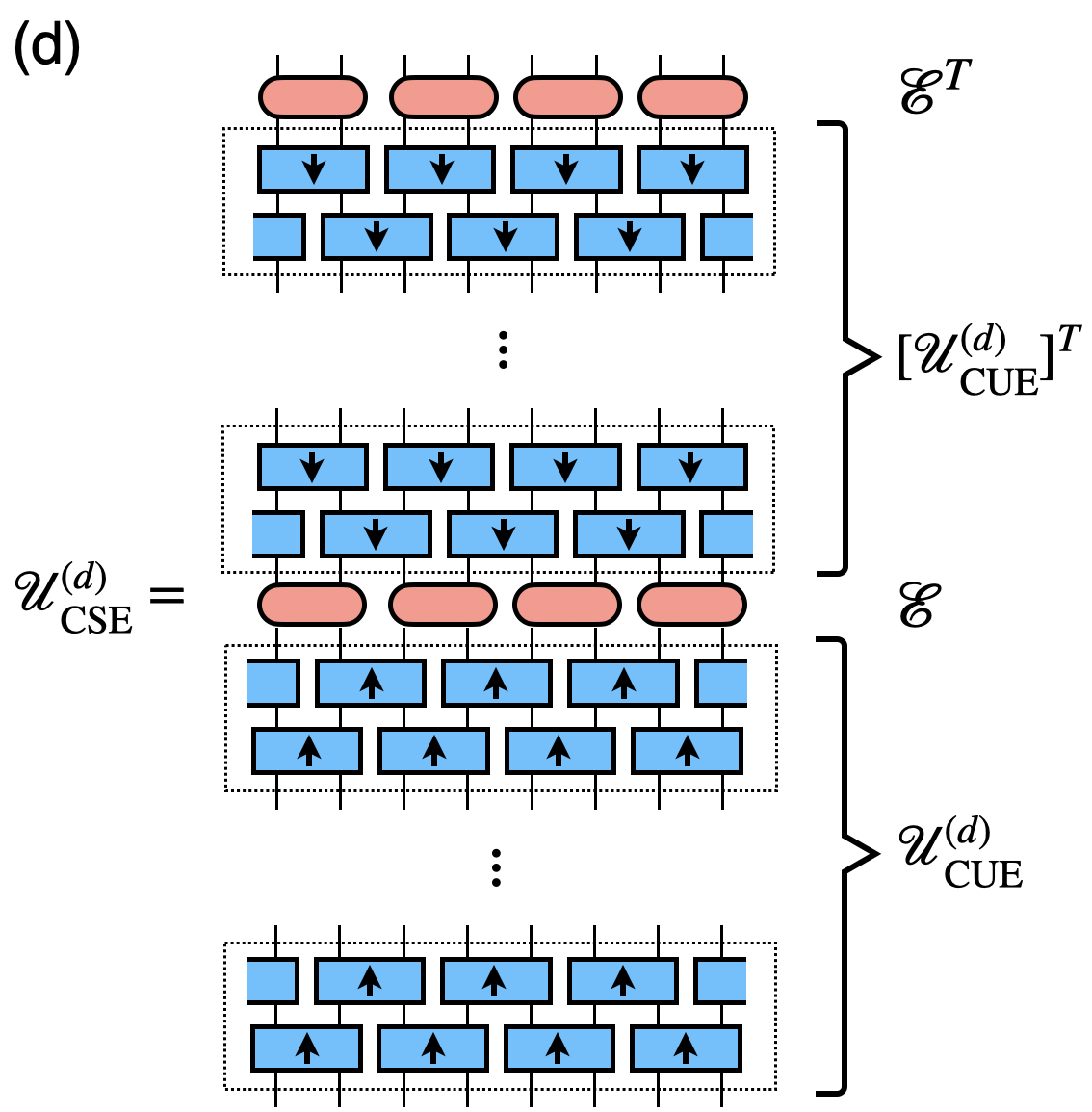}
    \caption{(a) $U_\mathrm{layer}$ consisting of nearest-neighbor matchgates. (b,c,d) Quantum circuit realizations of single-particle chaos corresponding to the (b) CUE, (c) COE, and (d) CSE.}
    \label{fig:circuits}
\end{figure*}

To realize the CUE $(\beta=1)$, we construct circuits out of independent random realizations of the gate $V_{i,i+1}$~\eqref{eq:Vmat}. A random realization is generated by drawing the phase $\phi_{0}$ from a uniform distribution over the range $[-\pi,\pi)$ and the $2\times 2$ matrix $u$ from the Haar measure over the 2-dimensional unitary group U$(2)$. The condition $\det A = \det B$ then uniquely determines the phase $\phi_{3}$. For simplicity, we assume $L_{\beta}$ is even; then a layer of the circuit is given by the brick layer unitary (see Fig.~\ref{fig:circuits})
\begin{align}\label{eq:Ulayer}
    U_\mathrm{layer} \equiv \prod_{m=1}^{{L_\beta}/2} V_{2m,2m+1} \prod_{m=1}^{{L_\beta}/2} V_{2m-1,2m},
\end{align}
where periodic boundary conditions are assumed. The one-cycle unitary is then formed by $d$ independently random iterations of $U_{\text{layer}}$,
\begin{align}\label{eq:U_CUE}
    \mU_\mathrm{CUE}^{(d)} \equiv \prod_{\tau=1}^d U_\mathrm{layer}^{(\tau)}.
\end{align}
Per the Solovay-Kitaev theorem and related results for unitary designs, we conjecture that this matchgate ensemble converges to the single-particle CUE in the limit $d\to\infty$~\cite{Solovay_Kitaev,NHJDesigns,NumberConservingDesigns}.

\begin{figure}
    \centering
    \includegraphics[width=\columnwidth]{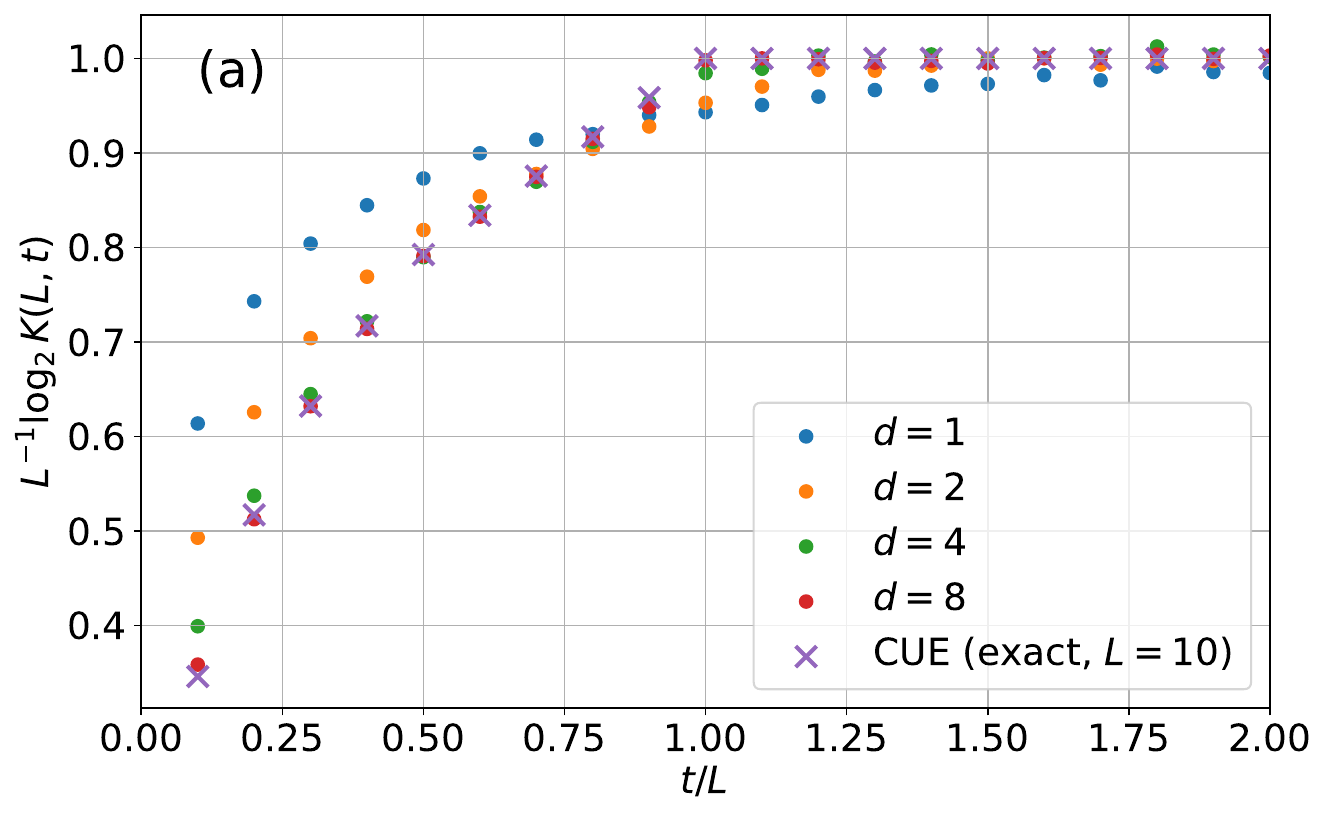}
        \includegraphics[width=\columnwidth]{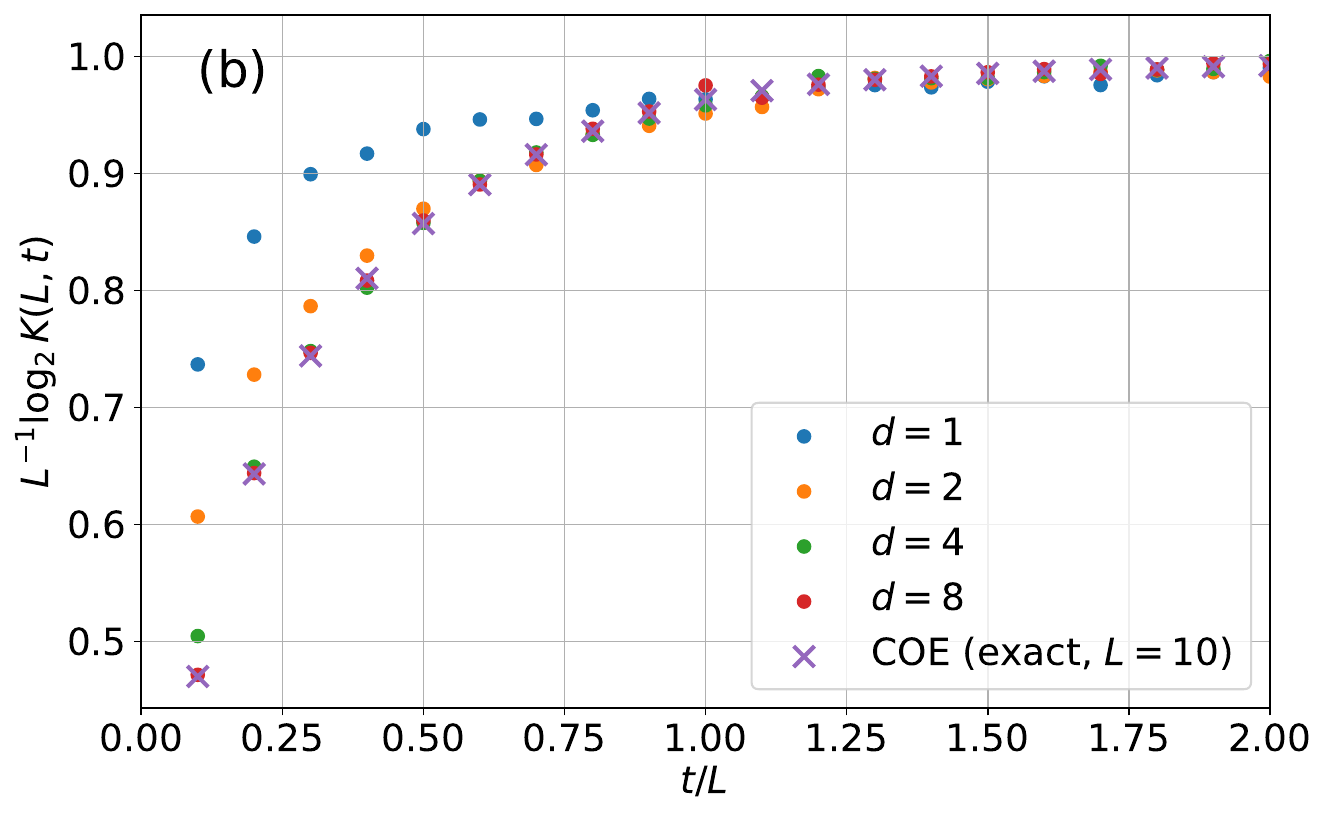}
    \includegraphics[width=\columnwidth]{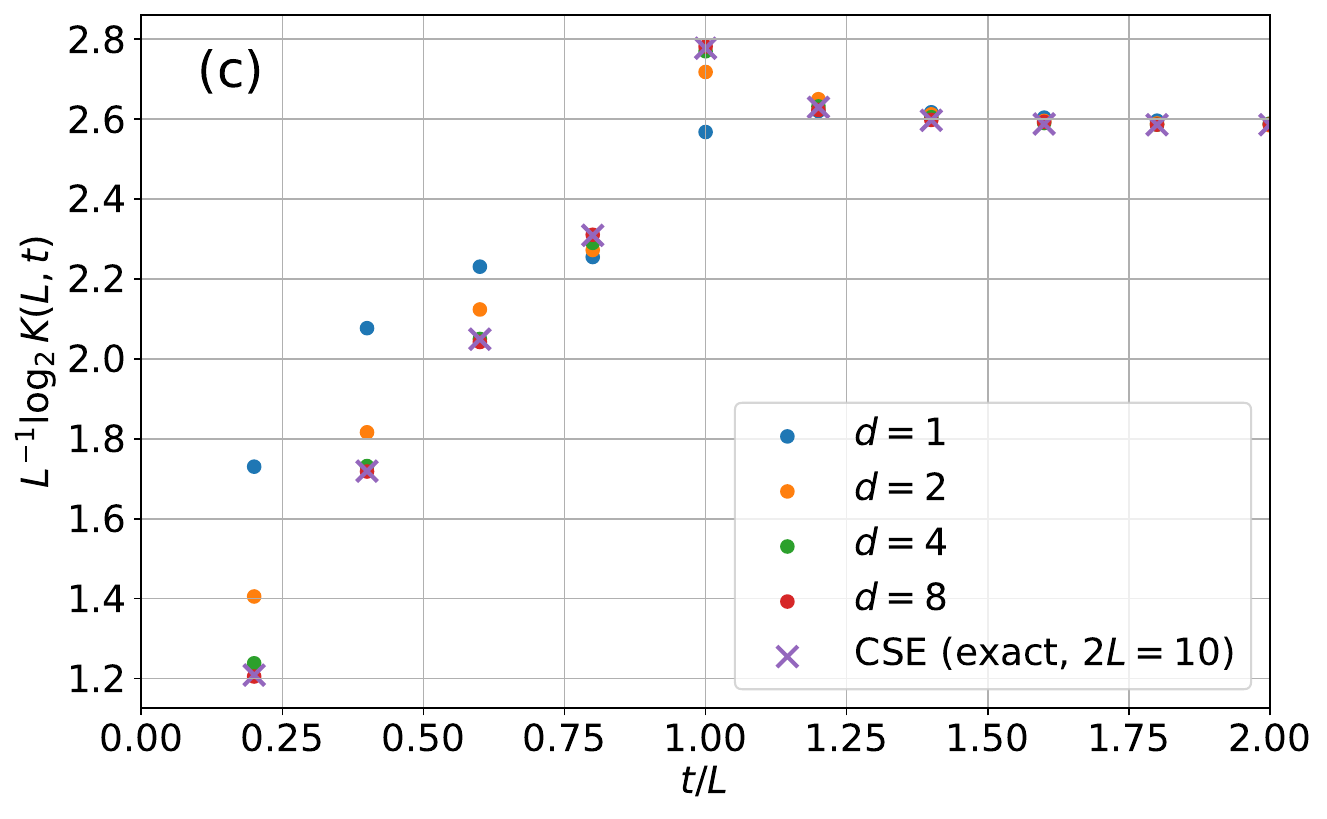}
    \caption{SFF of the depth-$d$ circuit with the nearest-neighbor matchgates corresponding to (a) the CUE ($\beta=1$), (b) COE ($\beta=2$), and (c) CSE ($\beta=4$). The numbers of qubits are ${L_\beta}=L=10$ in (a) and (b) and ${L_\beta}=2L=10$ in (c), and each data set for $d=1,2,4,8$ is obtained by averaging over $10^5$ samples. In each panel, crosses show the exact solution of the SFF for the CUE, COE, and CSE obtained in Sections~\ref{sec:CUE} and \ref{sec:CSEandCOE}.}
    \label{fig:SFF_gates}
\end{figure}

Figure~\ref{fig:SFF_gates}(a) shows the SFF numerically calculated by averaging the SFF of $\mU_\mathrm{CUE}^{(d)}$ over $10^5$ samples. As $d$ increases, the SFF approaches the exact solution obtained for the CUE model in Sec.~\ref{sec:CUE}. In contrast to circuits constructed from completely random two-qubit gates, our $U(1)$-conserving matchgates have the special structure of Eq.~\eqref{eq:U_and_H} and the underlying fermions are noninteracting. Thus, the $d\to\infty$ limit yields circuits which explore the space of non-interacting fermion systems whose single-particle states are described by CUE statistics.

Next, we consider the COE $(\beta=2)$, whose elements are symmetric unitaries of the form $U^T U$. It is then natural to use our matchgate representation of CUE unitaries to define the COE circuit
\begin{align}
    \mU_\mathrm{COE}^{(d)} = (\mU_\mathrm{CUE}^{(d)})^T \mU_\mathrm{CUE}^{(d)},
\end{align}
which is shown in Fig.~\ref{fig:circuits}(b). Arrows indicate the flow of time, so that the upper block is the transpose of the lower block. Figure~\ref{fig:SFF_gates}(b) shows the SFF obtained by numerical sampling of $\mU_\mathrm{COE}$ averaged over $10^5$ samples. As expected, it approaches the SFF for the COE as $d$ increases.

Finally, we consider the CSE ($\beta=4$) using a similar strategy. To conform to the notation in Sec.~\ref{sec:CSE}, we consider circuits with $L_4=2L$ qubits. The CSE is an ensemble of unitary matrices which are self-dual quaternions. Such unitaries can be written as $E^T U^T E U$, where $U$ is a CUE unitary and $E=\bigotimes_{m=1}^{L}(-i\sigma_y)_{2m-1,2m}$, where $\sigma_y$ is a Pauli matrix~\cite{mehta1991random}. We therefore define the following ensemble of circuits,
\begin{align}
    \mU_\mathrm{CSE}^{(d)}&=\mathcal{E}^T (\mU_\mathrm{CUE}^{(d)})^T \mathcal{E} \mU_\mathrm{CUE}^{(d)},\\
    \mathcal{E} &\equiv \prod_{m=1}^L W_{2m-1,2m},
\end{align}
where
\begin{align}\label{eq:Wmat}
    W_{i,i+1}=\begin{pmatrix}
1 & 0 & 0 & 0\\
0 & 0 & 1 & 0\\
0 & -1 & 0 & 0 \\
0 & 0 & 0 & 1
\end{pmatrix}
=\includegraphics[width=1.5cm,valign=c]{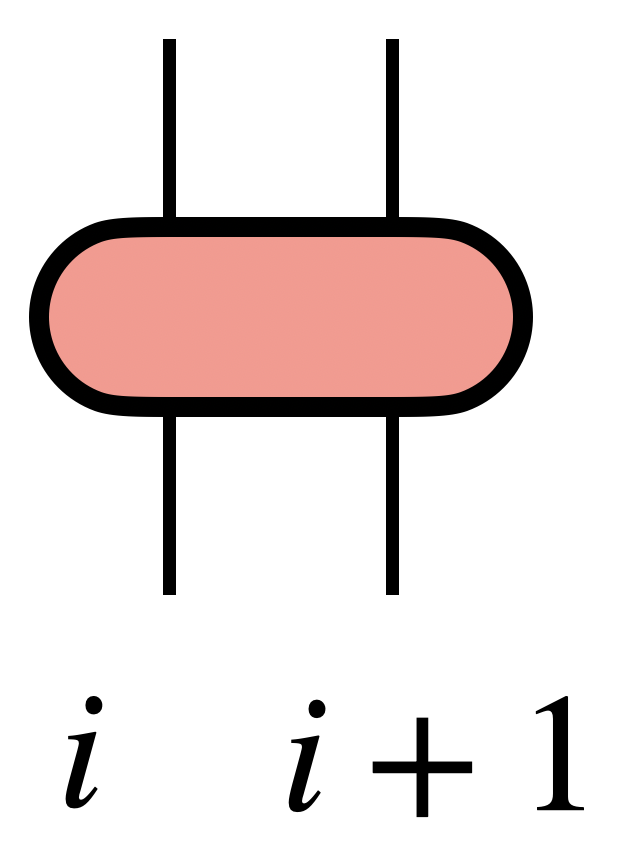}
\end{align}
is a non-random gate.
Note that $\mU_\mathrm{CSE}^{(d)}$ is defined as an $L_4=2L$-qubit circuit (see Fig.~\ref{fig:circuits}(c)).
Figure~\ref{fig:SFF_gates}(c) shows the SFF of this ensemble obtained by averaging over $10^5$ samples. Clearly, as we increase the depth $d$, the SFF approaches the CSE result. Thus, we have obtained circuit realizations of the SFF for single-particle random matrix statistics corresponding to each of the CUE, COE, and CSE.

Before closing this section, we remark that the SFF of a quantum circuit ensemble can be measured using the Hadamard test~\cite{Cleve1998}. This technique utilizes one ancillary qubit to control a matchgate and allows us to measure the real and imaginary parts of $\tr (\mU^t)$ for a given set of random parameters and various $t$, yielding the SFF after averaging over many samples. Also, the SFF can be estimated with randomized measurements without ancillary qubits~\cite{Joshi22,Dong2024}. Matchgate ensembles could thereby serve as an experimental realization of single-particle random matrix statistics in quantum simulators.

\section{Discussion and Conclusion}\label{sec:conclusion}

In this work, we have introduced methods which enable the exact computation of SFFs in a class of non-interacting models with spatially correlated disorder. While these models appear fine-tuned at first glance, they are, in fact, representatives of a generic class of models: non-interacting fermion systems with random Hamiltonian parameters. We explored this connection numerically in Sec.~\ref{sec:circuits} by sampling the SFFs of random free Hamiltonians and expect that this point of view will yield further insights in future work.

Our results provide a comprehensive understanding of
the spectral statistics in noninteracting many-body systems with Dyson statistics. For the single-particle CUE, we obtained an exact closed form for the SFF, valid for any system size $L$ and integer time $t$, without using approximations of any kind. The SFF exhibits a universal scaling collapse
and consists of a sequence of exponential ramps whose growth rates depend on the ratio $t/L$. For the single-particle COE and CSE, we developed numerically exact transfer matrix methods to compute their form factors. These approaches rely on a moment expansion, which expresses the SFF in terms of cycle contributions that can be evaluated without sampling.

A key finding of our work is that the exponential growth of the many-body SFFs can be understood as a consequence of the underlying single-particle random matrix statistics. This is in contrast to the linear ramp behavior typically associated with many-body quantum chaos, where random matrix universality emerges due to interactions. Our results thus highlight a novel mechanism for the onset of random matrix behavior in many-body systems, distinct from the usual interacting scenarios.

From a technical perspective, our work introduces several new tools for studying spectral statistics in many-body systems. The moment expansion and dimer embedding techniques developed for the CUE may find applications in related many-body problems, while the transfer matrix methods for the COE and CSE offer a powerful approach for systems where exact solutions are not available. Adapting these techniques to more general settings is an exciting avenue for future research.

To bridge the gap between our non-interacting models and realistic chaotic many-body systems, we constructed ensembles of matchgate quantum circuits that mimic the single-particle circular ensembles. These provide a practical platform for investigating the crossover from single-particle to many-body random matrix statistics and open the door to experimental realizations in quantum simulators. An interesting future direction is to study the emergence of the many-body linear ramp by gradually introducing interactions into these matchgate circuits.

Our results also provide a baseline to analyze the emergence of many-body random matrix universality from single-particle states which exhibit random matrix statistics themselves. A number of well-known models can be used to study this crossover, such as the kicked Ising model with one-step evolution
\begin{equation}
   \mU_\mathrm{KIM} = \exp\left[ -i  \sum_{i=1}^L \left(J Z_i Z_{i+1}+\frac{\theta_i}{2} Z_i \right)\right] \exp\left[-i g \sum_{i=1}^L X_i \right]
\end{equation}
where the $\theta_i$ are CUE quasienergies and $Z_i, X_i$ are Pauli matrices. The authors initiated study of the single-particle/many-body crossover in this model in a companion paper~\cite{OurPaper1}, but this study is far from exhaustive. We anticipate that further developments in this direction are a rich avenue for future research.

\section*{Acknowledgments}
We acknowledge fruitful discussions with Thomas Scaffidi, Ceren Dag, Keisuke Fujii, and Kaoru Mizuta.
T. N. I. was supported by JST PRESTO Grant No.~JPMJPR2112, JSPS KAKENHI Grant No.~JP21K13852, and the Boston University CMT visitors program.
L.V. acknowledges support from the Slovenian Research and Innovation Agency (ARIS), Research core funding Grants No.~P1-0044, N1-0273, J1-50005 and N1-0369, as well as the Consolidator Grant Boundary-101126364 of the
European Research Council (ERC). M.O.F. acknowledges support from the Faculty of Science at the University of Victoria through Thomas E. Baker. This research was supported in part by the International Centre for Theoretical Sciences (ICTS) for participating in the program -- Stability of Quantum Matter in and out of Equilibrium at Various Scales (code:ICTS/SQMVS2024/01).

\bibliography{Paper2Refs}
\end{document}